\documentclass{article}
\usepackage[english]{babel}

\usepackage{theorem}
\usepackage{latexsym}
\usepackage{amsfonts}
\usepackage{amssymb}
\usepackage{multicol}

\newenvironment{pf}{\noindent{\it Proof. }}{\fbox{}\\}
\newtheorem{thm}{Theorem}
\newtheorem{prop}{Proposition}
\newtheorem{lem}{Lemma}
\newtheorem{cor}{Corollary}
{\theorembodyfont {\rmfamily}
\newtheorem{rem}{Remark}
\newtheorem{defi}{Definition}
}

\begin{document}


\title{\bf{Large deviations for quantum Markov semigroups on the  $2\times 2$-matrix algebra}}

\author{Henri Comman\thanks{Department of Mathematics, University
of Santiago de Chile, Bernardo O'Higgins 3363, Santiago, Chile.
E-mail: hcomman@mat.usach.cl}}

\date{}

\maketitle

\abstract{Let $({\mathcal{T}}_{*t})$ be a  predual quantum Markov
semigroup
 acting on the full $2\times 2$-matrix
algebra and having an absorbing pure state. We prove that for any
initial state $\omega$, the net of orthogonal measures representing
the net of states $({\mathcal{T}}_{*t}(\omega))$ satisfies a large
deviation principle in the pure state space, with a  rate function
given in terms of the generator, and which does not depend on
$\omega$. This implies that $({\mathcal{T}}_{*t}(\omega))$ is
faithful for all $t$ large enough. Examples arising in weak coupling
limit are studied.}

\section{Introduction}

 The integral representation of states on a unital separable
$C^*$-algebra establishes that each  state is the barycentre of a
measure concentrated on the set of pure states $P$ (\cite{tak}).
There are in general various such representing measures, a class of
which is the so called orthogonal measures. In the case of the
algebra of compact operators on some separable complex Hilbert space
$\mathcal{H}$, when the state $\omega$ is given by the positive
trace-one operator $\rho$,  to each diagonal form
$\rho=\sum_{i=1}^{\infty}a_i|e_i\rangle\langle e_i|$ is associated
the  orthogonal measure
$\mu=\sum_{i=1}^{\infty}a_i\delta_{\omega_{|e_i\rangle\langle
e_i|}}$, where $\omega_{|e_i\rangle\langle e_i|}$ is the pure state
given by the projection  $|e_i\rangle\langle e_i|$ (note that such a
measure is  uniquely determined by $\rho$ if and only if  all
eigenvalues are simple).

In this paper we study large deviations  for nets of orthogonal
measures, and in particular  when these nets are  given by a quantum
Markov semigroup acting on the full $2\times 2$-matrix algebra
$M_2$.
 More precisely, let
 $({\mathcal{T}}_t)$  be  such a semigroup having
   an absorbing
  state $\omega_\infty$ (i.e., in  physical terminology, $({\mathcal{T}}_t)$ converges
 to the
 equilibrium), and let $({\mathcal{T}}_{*t})$ denotes its predual
 semigroup.
  For each initial state $\omega$, we consider the net of
 states $({\mathcal{T}}_{*t}(\omega))$.
 Our main result establishes that when  $\omega_\infty$ is pure,
 the   net of orthogonal measures representing
 $({\mathcal{T}}_{*t}(\omega))$
satisfies a large deviation principle in $P$ with powers $(1/t)$;
 the rate
function  takes the  values $\{0,\eta-a,+\infty\}$ where $a,\eta$
 are  parameters given by  the generator of $({\mathcal{T}}_{*t})$, and in particular
 it does not
depend on $\omega$ (Theorem \ref{finite-case}). This gives an
exponential rate  of  "purification" of the state
${\mathcal{T}}_{*t}(\omega)$
 in terms of the generator, in the large
 deviation sense (i.e. the  rate with which the mass assigned to
 sets not containing the limit state vanishes). This rate  is given by
 the
 eigenvalues of the operator $\mathcal{J}^*(|e_1\rangle\langle e_1|)$,
 where $e_1$ is the unit vector
  determining $\omega_\infty$,
  and
 $\mathcal{J}$
 a completely positive operator on $M_2$ appearing in the
 generator. As a consequence, we obtain an exponential rate of convergence of
the semigroup on projections (Corollary \ref{NC-LDP}); this result
can be interpreted as a  noncommutative large deviation principle as
defined by the author in previous works, and  the so-called "rate
operator" is  exactly $\mathcal{J}^*(|e_1\rangle\langle e_1|)$ (see
Remark \ref{remark-NC-LDP}).

The proof rests essentially on two operator-theoretic ingredients:
the first one is the well-known  form of the generator of quantum
Markov semigroups acting on all bounded operators on $\mathcal{H}$,
and having a pure stationary state; the second one is a general
result that we prove for these semigroups, when they act on $M_2$
and admit an absorbing state. It establishes that
${\mathcal{T}}_{*t}(\omega)$ is faithful for all states $\omega\neq\omega_\infty$ and
all $t$ large enough (Theorem \ref{geom-prop}); this property will
be, in turn, recovered as a consequence of large deviations. By a
compactness argument combined with large deviations techniques, this
result   allows us to reduce the proof of the general initial state
case to the one given by $\frac{1}{2}I$, where $I$ is the identity.

Although we are mainly interested by orthogonal measures arising
from dynamics as above, we begin in section \ref{section-LDP} by
considering a general family of such measures ($\mathcal{H}$
infinite dimensional), for which we give sufficient conditions to
have large deviations (Proposition \ref{prop-conditions}). This
requires recent results in large deviation theory,  and in
particular a notion of exponential $\tau$-smoothness, weaker than
the usual exponential tightness (\cite{com2}).
 We then specialize  to the case where $\mathcal{H}$ is $N$-dimensional and
 the net of states is
  converging (Proposition \ref{LDP-converging}).

The problem of large deviations for orthogonal measures given by the
evolution of  quantum Markov semigroups is posed even  in absence of
convergence; this  contrasts with  the usual approach where it is
 the distance to some  limit state
 which is measured. As a  motivation to study in this way the asymptotic behavior
  we can mention some models of
 information dynamics, where algorithms are represented by the
 semigroups, the input by the initial state and the output by the limit state, which is
typically pure (\cite{Belavkin-Ohya-02-Roy-Soc-Lon-Proc}); the
complexity of the state under the evolution is represented by its
support.  Indeed, our large deviations describe the rate with which
this support decreases.

 In fact, the
method used here for the two-dimensional case gives some indications
about
 possible  extensions to higher  dimensions.
Since the main tools for the proofs
 are the representation of the generator given by Theorem
 \ref{Davies-result} and large deviations techniques, which
 both are valid in higher dimensions,
  we could reasonably expect that similar results
 hold at least in finite dimensions when there exists an absorbing pure state (the infinite
  dimensional case is  more delicate
   because of the non-compactness of the pure state space).
A crucial argument in the proof uses the fact that the operator $y$
in (\ref{Davies-result-eq1}) is diagonalizable, which is a
particular feature of dimension two; this suggests that in dimension
$N$,
 some extra conditions on the generator may have to be add.
Note that Proposition \ref{LDP-converging} shows that a strict
$N$-dimensional analogue of  the large deviation principle proved
here would imply the convergence of eigenvectors. It is likely that
such a large deviation principle implies its noncommutative
counterpart, namely an exponential rate of convergence on
projections (see Remark \ref{remark-NC-LDP}); in other words, part
$(b)$ of Corollary \ref{NC-LDP} should admit a generalization. Since
any noncommutative large deviation principle  admits a unique rate
operator (\cite{comman-JOT-06}, Proposition 5.2),  a natural
question arises: Is this rate operator
 still given by $\mathcal{J}^*(|e_1\rangle\langle e_1|)$?
 Or coming back to the classical setting: Do the
 eigenvalues of $\mathcal{J}^*(|e_1\rangle\langle e_1|)$ still
 correspond to the finite values of the rate function?
On the other hand, the only hypothesis in Theorem
 \ref{Davies-result} being the existence of a stationary pure state, a
 similar study can be made in  more general situations when there is no
  absorbing state. For
 instance,
  when there is a  set $S_\infty$ of pure states
   such that for  each initial state
  $\omega$,
   $\mathcal{T}_{*t}(\omega)$
  converges to some  element of $S_\infty$;
  we should then obtain various  rate
   functions indexed by the elements of $S_\infty$.
  This kind of semigroups belongs to the  class of the so-called
   "generic" semigroups, which
     arise in an extended version of the weak coupling
   limit (\cite{Acc-Fa-Ha-06-IDAQP}).
 A class of such semigroups
  admitting an  absorbing pure state (i.e. $S_\infty$ contains only  one
  element) is studied in Section \ref{section-ex}.

\subsection{Notations and background material}

\subsubsection{Quantum Markov semigroups}

Let $\mathcal{H}$ be a  complex separable Hilbert space, and let
$K(\mathcal{H})$ be the set of compact operators acting on
$\mathcal{H}$.
 Let $P$ be the  pure state
space of $K(\mathcal{H})$ provided with the weak$^*$ topology, and
note that $P$ is  completely regular Hausdorff, and compact when
$\mathcal{H}$ is finite dimensional. For each $x\in K(\mathcal{H})$,
we denote by $\hat{x}$ the map defined on $P$ by
$\hat{x}(\omega)=\omega(x)$, and note that $\hat{x}$ is continuous.
The full $N\times N$-matrix algebra is denoted by $M_N$. By
convention, for any self-adjoint $\rho\in M_N$, the expression
$\rho=\sum_{i=1}^N a_i|e_i\rangle\langle e_i|$ has to be considered
as a formal sum (i.e., $a_i$ can be zero), which means that the set
$\{e_i:1\le i\le N\}$ is an orthonormal  basis diagonalizing $\rho$,
where each $e_i$ is an eigenvector corresponding to the eigenvalue
$a_i$.

Let $\omega_\rho$ denotes the state given by the positive trace-one
operator  $\rho$. When $\rho$ admits a diagonal form
$\rho=\sum_{i=1}^{\infty}a_i|e_i\rangle\langle e_i|$, we will
consider the measure
$\mu=\sum_{i=1}^{\infty}a_i\delta_{\omega_{|e_i\rangle\langle
e_i|}}$, where the sum has to be understood in the sense of the weak
topology for Borel measures on $P$. It is easy to see that $\mu$ is
an orthogonal measure  representing $\omega_\rho$, in the sense of
the theory of integral representation of states (\cite{tak}).
Clearly, when $ \mathcal{H}$ has dimension $2$, each state distinct of $I/2$ is
represented by a unique orthogonal measure. We shall use the
following lemma whose proof is straightforward.

\begin{lem}\label{conv-meas-conv-states}
For any net $(\omega_t)$ of states on $M_N$ and any state $\omega$
on $M_N$, the following statements are equivalent.
\begin{itemize}
\item[(i)] $\lim \omega_t=\omega$;
\item[(ii)] $\omega$ (resp. $\omega_t$) is represented by an
 orthogonal measure $\mu$ (resp. $\mu_t$) such that $\lim\mu_t (\hat{x})=\mu(\hat{x})$ for all $x\in M_N$;
\item[(iii)]  $\lim\mu_t (\hat{x})=\mu(\hat{x})$ ($x\in M_N$) for all orthogonal measures
$\mu_t$ and $\mu$ representing $\omega_t$ and $\omega$,
respectively.
\end{itemize}
\end{lem}

By a quantum Markov semigroup acting on $M_2$, we mean a
$w^*$-continuous one-parameter semigroup $({\mathcal{T}}_t)_{t\ge
0}$ of completely positive linear maps on $M_2$ preserving the
identity $I$. We denote by $({\mathcal{T}}_{*t})$ the predual
semigroup, and by $(\widetilde{\mathcal{T}}_{*t})$ the associated
semigroup acting on $M_2$, obtained by identifying $\omega_\rho$ and
$\rho$. In other words, $(\widetilde{\mathcal{T}}_{*t})$ is a
strongly continuous semigroup of completely positive contractions on
$M_2$ defined by the relation
$\omega_{\widetilde{\mathcal{T}}_{*t}(\rho)}=\mathcal{T}_{*t}(\omega_{\rho})$
for all states $\omega_\rho$ and extended  by linearity. A state
$\omega$ is stationary if ${\mathcal{T}}_{*t}(\omega)=\omega$ for
all $t\ge 0$, and a state $\omega_\infty$ is said to be absorbing if
$\lim{\mathcal{T}}_{*t}(\omega)=\omega_\infty$ for all states
$\omega$; note that an absorbing state is stationary.

In the following theorem, we collect various results that  will be
used in the sequel. They appear in \cite{dav}, and are  valid more
generally for  uniformly continuous  quantum Markov semigroups
acting on the algebra of all bounded operators on $\mathcal{H}$
infinite dimensional.

\begin{thm}\label{Davies-result}
Let $({\mathcal{T}}_t)$ be a quantum Markov semigroup on $M_2$
having a stationary pure state given by the projection
$|e\rangle\langle e|$. Then there exist $y,z_1,z_2$ in $M_2$ such
that the following hold.
\begin{itemize}
\item[(a)]
 The generator
$\tilde\mathcal{L}_*$ of $(\tilde{\mathcal{T}}_{*t})$ has the form
\begin{equation}\label{Davies-result-eq1}
\forall \rho\in M_2,\ \ \ \ \ \ \tilde\mathcal{L}_*(\rho)=y\rho+\rho
y^*+\mathcal{J}(\rho),
\end{equation}
where $\mathcal{J}$ is defined on $M_2$ by
$\mathcal{J}(\rho)=\sum_{i=1}^2z_i\rho z_i^*$.
\item[(b)] $ye=y^*e=z_1e=z_2e=0$.
\item[(c)] $y$ is the generator of a one-parameter semigroup of contractions $(C_t)_{t\ge 0}$ on $\mathcal{H}$,
and the semigroup $(\mathcal{S}_t)_{t\ge 0}$ on $M_2$ defined by
$\mathcal{S}_t(\rho)=C_t\rho C_t^*$, satisfies for all $t\ge 0$,
\begin{equation}\label{Davies-result-eq3}
\forall \rho\ge 0,\ \ \ \ \
0\le\mathcal{S}_t(\rho)\le\tilde\mathcal{T}_{*t}(\rho)
\end{equation}
and
\begin{equation}\label{Davies-result-eq4}
\tilde\mathcal{T}_{*t}=\mathcal{S}_t+\int_0^t\tilde\mathcal{T}_{*t-s}\mathcal{J}\mathcal{S}_s
ds.
\end{equation}
\end{itemize}
\end{thm}

\subsubsection{Large deviations}\label{subsection-LDP}

We recall now some  large deviations  results for a  net
$(\mu_t)_{t\ge 0}$ of Borel probability measures on a completely
regular  Hausdorff topological space $X$. For each
$[-\infty,+\infty[$-valued Borel measurable function $h$ on $X$, we
put $\mu_t^{1/t}(e^{th})=(\int_X e^{th(x)}\mu_t(dx))^{1/t}$, and
  define $\Lambda(h)=\log\lim
\mu_t^{1/t}(e^{th})$ when the limit exists. By definition, $(\mu_t)$
satisfies a large deviation principle with powers $(1/t)$ if there
exists a $[0,+\infty]$-valued lower semi-continuous function $J$ on
$X$ such that
\[
\limsup\mu_t^{1/t}(F)\le\sup_{x\in F}e^{-J(x)}\ \ \ \ \ \ \
\textnormal{for all closed $F\subset X$}
\]
and
\[
\sup_{x\in G}e^{-J(x)}\le
 \liminf\mu_t^{1/t}(G)\ \ \
\ \ \ \ \textnormal{for all open $G\subset X$};
\]
$J$  is called the  rate function for $(\mu_t^{1/t})$, and for each
Borel set
 $A\subset X$  such that $\sup_{x\in
\textnormal{Int}(A)}e^{-J(x)}=\sup_{x\in \overline{A}}e^{-J(x)}$,
 the limit
$\lim\mu_t^{1/t}(A)$ exists and satisfies
\begin{equation}\label{subsection-LDP-eq1}
\lim\mu_t^{1/t}(A)=\sup_{x\in A}e^{-J(x)}
\end{equation} (such a
set $A$ is called a $J$-continuity set). The following notions have
been introduced in \cite{com2}.

\begin{defi}\label{tau-smooth}
The net $(\mu_t)$ is exponentially   \textit{$\tau$-smooth} if for
all open covers $\{G_i:i\in I\}$ of $X$ and for all $\varepsilon>0$,
 there exists a finite set $\{G_{i_1},...,G_{i_N}\}\subset\{G_i:i\in I\}$ such that
 \[
\limsup\mu_t^{1/t}(X\verb'\'\bigcup_{1\le j\le N}{G_{i_j}})
 <\varepsilon.
\]
\end{defi}

\begin{defi}\label{defi-approx-class}
A class $\mathcal{A}$ of $[-\infty,+\infty[$-valued continuous
functions on $X$
 is said to be  \textit{approximating} if for each $x\in X$,  each
 open set
$G$ containing $x$, each real $s>0$, and each real $r>0$,
$\mathcal{A}$ contains some function $h$ satisfying
\[
e^{-r}1_{\{x\}}\le e^{h}\le 1_G\vee e^{-s}.
\]
\end{defi}

Under exponential $\tau$-smoothness, the existence of
$\Lambda(\cdot)$ on some approximating class $\mathcal{A}$ is
sufficient to get large deviations, with a rate function which can
be expressed in terms of $\mathcal{A}$. However, with slight extra
conditions on $\mathcal{A}$, this expression is substantially
simplified (\cite{com2}, Corollary 2 and Corollary 4). This variant
is stated in the following theorem, and  will be used in the proof
of Proposition \ref{prop-conditions}.

\begin{thm}\label{funct-LDP-cond}
Let $\mathcal{A}$ be an approximating class of bounded above
functions such that for each $x\in X$,  each
 open set
$G$ containing $x$, and  each real $s>0$,  $\mathcal{A}$ contains
some function $h$ satisfying
\[
1_{\{x\}}\le e^{h}\le 1_G\vee e^{-s}.
\]
  If $(\mu_t)$ is exponentially $\tau$-smooth
and if $\Lambda(h)$ exists for all $h\in\mathcal{A}$, then $(\mu_t)$
satisfies a large deviation principle with powers $(1/t)$ and rate
function
\[
J(x)=\sup_{h\in\mathcal{A},h(x)=0}\{-\Lambda(h)\}\ \ \ \ \ \ for \
all\  x\in X.\]
\end{thm}

\section{Large deviations for orthogonal measures}\label{section-LDP}

In this section, we first give  sufficient conditions for a net of
orthogonal measures to satisfy a large deviation principle
(Proposition \ref{prop-conditions}). Next, we specialize to the case
where $\mathcal{H}$ is finite dimensional and the net of states is
converging (Proposition \ref{LDP-converging}).

\begin{prop}\label{prop-conditions}
Let $(\omega_t)_{t\ge 0}$ be a net of states on $K(\mathcal{H})$,
where each $\omega_t$ is represented by the orthogonal  measure
$\mu_t=\sum_{i=0}^{\infty}a_{i,t}\delta_{\omega_{|e_{i,t}\rangle\langle
e_{i,t}|}}$, and assume that the following conditions hold:
\begin{itemize}
\item[(i)]
$\lim_N\limsup(\sum_{i=N}^{\infty}a_{i,t})^{1/t}=0$.
\item[(ii)] The net
 $(\omega_{|e_{i,t}\rangle\langle e_{i,t}|})_{t\ge 0}$ converges
 in $P$,  for all $i\in\mathbb{N}$.
\item[(iii)]
 $\lim a_{i,t}^{1/t}$
 exists for all $i\in\mathbb{N}$.
\end{itemize}
Then, $(\mu_t)$ satisfies a large deviation principle  with powers
$(1/t)$ and rate function
\begin{equation}\label{prop-conditions-eq0}
J(\omega)=\sup_{h\in\mathcal{A},h(\omega)=0}\{-\Lambda(h)\}\ \ \ \ \
\textnormal{for all}\ \omega\in P,\end{equation} where
$\mathcal{A}=\bigcup_{\omega\in P}\mathcal{A}_\omega$ with
$\mathcal{A}_\omega$ the set of all finite infima of elements in
$\{-|\hat{x}-\hat{x}(\omega)|:x\in K(\mathcal{H})\}$.
\end{prop}

\begin{pf}
  By $(ii)$, for each $i\in\mathbb{N}$ there exists $e_i\in\mathcal{H}$ such that
   $\lim\omega_{|e_{i,t}\rangle\langle e_{i,t}|}=\omega_{|e_i\rangle\langle
e_i|}$. Let $\mathcal{G}_0$ be an open cover of $P$, and let for
each $i\in\mathbb{N}$ some $G_i\in\mathcal{G}_0$ containing
$\omega_{|e_i\rangle\langle e_i|}$.
 By $(i)$, for each  $\varepsilon>0$ there exists
$N_\varepsilon\in\mathbb{N}$ such that
$\limsup(\sum_{i=N_\varepsilon+1}^{\infty}a_{i,t})^{1/t}<\varepsilon$.
Since for each $t\ge 0$,
\[\mu_t(P\verb'\'\cup_{i=0}^{N_\varepsilon}G_i)=
\sum_{i=0}^{N_\varepsilon}
a_{i,t}\delta_{\omega_{|e_{i,t}\rangle\langle
e_{i,t}|}}(P\verb'\'\cup_{i=0}^{N_\varepsilon}G_i)+\sum_{i=N_\varepsilon+1}^{\infty}a_{i,t}\delta_{\omega_{|e_{i,t}\rangle\langle
e_{i,t}|}}(P\verb'\'\cup_{i=0}^{N_\varepsilon}G_i),\] with $G_i$
containing  $\omega_{|e_{i,t}\rangle\langle e_{i,t}|}$ for all
$i\in\{0,...,N_\varepsilon\}$ and all $t$ large enough, we get
\[\limsup\mu_t^{1/t}(P\verb'\'\cup_{i=0}^{N_\varepsilon}G_i)<\varepsilon.\]
This shows that  $(\mu_t)$ is exponentially $\tau$-smooth since
$\mathcal{G}_0$ is arbitrary.
 For each
$\omega\in P$, each  open set $G\subset P$ containing $\omega$, and
each $s>0$, by definition of the $w^*$-topology, there exists a
finite set $K_{\omega,G,s}\subset K(\mathcal{H})$ such that
\[1_{\{\omega\}}\le e^{-\sup_{x\in K_{\omega,G,s}}|\hat{x}-\hat{x}(\omega)|}\le 1_G\vee
e^{-s},\]hence  $\mathcal{A}$ is an approximating class for $P$ satisfying the hypothesis of Theorem \ref{funct-LDP-cond}.
 For
any $\omega\in P$ and any finite subset $K\subset K(\mathcal{H})$,
there is some $N\in\mathbb{N}$ such that
\[\limsup\mu_t^{1/t}(e^{-t\sup_{x\in K}
|\hat{x}-\hat{x}(\omega)|})= \limsup(\sum_{i=1}^N
a_{i,t}e^{-t\sup_{x\in K}|\langle e_{i,t},x
e_{i,t}\rangle-\omega(x)|})^{1/t}\]
\begin{equation}\label{prop-conditions-eq1}
=\sup_{1\le i\le N} \lim a_{i,t}^{1/t}e^{- \sup_{x\in K}|\langle
e_{i,t},x e_{i,t}\rangle-\omega(x)|}=\lim\mu_t^{1/t}(e^{-t\sup_{x\in
K}|\hat{x}-\hat{x}(\omega)|}),
\end{equation} where the first
equality follows from $(i)$ and the second from $(ii)$ and $(iii)$.
Since $\omega$ and $K$ are arbitrary, $\lim\mu_t^{1/t}(e^{th})$
exists and so $\Lambda(h)$ exists for all $h\in\mathcal{A}$. By
Theorem \ref{funct-LDP-cond},  $(\mu_t)$ satisfies a large deviation
principle with powers $(1/t)$ and rate function
(\ref{prop-conditions-eq0}).
\end{pf}

\begin{lem}\label{prop-eigen}
Let $(\rho_t)_{t\ge 0}$ be a   net of hermitian matrices in $M_N$
  converging in norm
 to a hermitian matrix  $\rho=\sum_{i=1}^N a_i|e_i\rangle\langle
e_i|$ with $a_{1}>...>a_{N}$. Then for each $t$ large enough
$\rho_t$  admits a diagonal form
 $\rho_t=\sum_{i=1}^N a_{i,t}|e_{i,t}\rangle\langle e_{i,t}|$ such
that  $\lim a_{i,t}=a_i$  and $\lim |e_{i,t}\rangle\langle
e_{i,t}|=|e_i\rangle\langle e_i|$ (in norm)  for all
$i\in\{1,...,N\}$.
\end{lem}

\begin{pf}
Let $m_{i,t}$ denote the multiplicity of   $a_{i,t}$ ($1\le i\le
N$).  Since $\rho_t$ converges in trace-norm, the assertion
concerning the eigenvalues follows from the well-known
 inequality $\sum_{i=1}^N|a_i-a_{i,t}|\le||\rho-\rho_t||_1$ where $||\cdot||_1$ denotes the trace norm (\cite{pow}).  Let
$\varepsilon<\frac{1}{2}\min_{i=1}^{N-1}\{|a_i-a_{i+1}|\}$.
 Since
$(\rho_t)$ is uniformly converging, it converges in norm resolvent
sense, so that $E^{\rho_t}_{]a_i-\varepsilon,a_i+\varepsilon[}$
converges uniformly  to
$E^{\rho}_{]a_i-\varepsilon,a_i+\varepsilon[}=E^{\rho}_{\{a_i\}}$
for each $i\in\{1,...,N\}$. For each $\varepsilon'\le\varepsilon/4$
 we have
\[
E^{\rho}_{\{a_1\}}=E^{\rho}_{]a_1-\varepsilon,a_1+\varepsilon[}=\lim_t
E^{\rho_t}_{]a_1-\varepsilon,a_1+\varepsilon[}\sum_{j=1}^N\frac{1}{m_{j,t}}
E^{\rho_t}_{]a_{j,t}-\varepsilon', a_{j,t}+\varepsilon'[}\]
\[=\lim_t \sum_{j=1}^N\frac{1}{m_{j,t}}
E^{\rho_t}_{]a_1-\varepsilon,a_1+\varepsilon[\cap]a_{j,t}-\varepsilon',
a_{j,t}+\varepsilon'[}=\lim_t \frac{1}{m_{1,t}}\sum_{j=1}^{m_{1,t}}
E^{\rho_t}_{]a_1-\varepsilon,a_1+\varepsilon[\cap]a_{j,t}-\varepsilon',
a_{j,t}+\varepsilon'[}\]
\[=\lim_t \frac{1}{m_{1,t}}\sum_{j=1}^{m_{1,t}}
E^{\rho_t}_{]a_{j,t}-\varepsilon', a_{j,t}+\varepsilon'[}=\lim_t
E^{\rho_t}_{]a_{1,t}-\varepsilon', a_{1,t}+\varepsilon'[}=\lim_t
E^{\rho_t}_{\{a_{1,t}\}}.
\]
Similarly  we get $\lim_t
E^{\rho_t}_{\{a_{i,t}\}}=E^{\rho}_{\{a_i\}}$ for all
$i\in\{2,...,N\}$, which proves the lemma.
\end{pf}

Part $(a)$ of the following proposition shows that when
$(\omega_{\rho_t})$ converges to some state $\omega_{\rho_\infty}$,
and under some extra condition on eigenvectors, large deviations for
a suitable representing net of orthogonal
 measures
are determined by the asymptotic behavior  of the eigenvalues of
$\rho_t$. The interesting  case occurs when $\omega_{\rho_\infty}$
is not faithful, otherwise the rate function
(\ref{LDP-converging-eq0}) is trivial since $r=N$; it  gives then
the  rate with which the support of $\rho_t$ gets smaller. Note that
by Lemma \ref{prop-eigen} the hypotheses are always satisfied in
dimension $2$ when $\omega_{\rho_\infty}\neq \frac{1}{2}I$, and in
particular when $\omega_\infty$ is pure. Although this will  not be
used in the sequel, it is worth noticing that (assuming $\lim
\omega_{\rho_t}=\rho_\infty$) a large deviation principle with
 rate function (\ref{LDP-converging-eq0})  implies the
convergence of some eigenvectors, as establishes $(b)$.

\begin{prop}\label{LDP-converging}
Let  $(\omega_{\rho_t})_{t\ge 0}$ be a net of states on $M_N$, and
assume that $(\omega_{\rho_t})$ $w^*$-converges to some  state
$\omega_{\rho_\infty}$. Let $a_{1,t}\ge...\ge a_{N,t}$ be the
eigenvalues of  $\rho_t$, and $a_1\ge...\ge a_r$ be the non-zero
eigenvalues of $\rho_\infty$ counted with multiplicity  ($a_i=0$ for
$i>r$), and consider the following property $(P_i)$ for any
$i\in\{1,...,N\}$.
\begin{itemize}
\item[$(P_i)$] For each $t$ large enough $a_{i,t}$ (resp. $a_i$) admits an eigenvector
$e_{i,t}$ (resp. $e_i$) such that
$\lim\omega_{|e_{i,t}\rangle\langle
e_{i,t}|}=\omega_{|e_i\rangle\langle e_i|}$.
\end{itemize}
Then,
\begin{itemize}
\item[(a)] If $(P_i)$ holds for all $i\in\{1,...,N\}$,
  then the associated net $(\mu_t)$  of orthogonal
 measures  satisfies a large
deviation principle
 with powers
$(1/t)$  if and only if
 $\lim \frac{1}{t}\log a_{i,t}$ exists for all $i\in\{r+1,...,N\}$. In this case, the  rate
 function is
  \begin{equation}\label{LDP-converging-eq0}
J(\omega_{|e\rangle\langle e|})=\left\{
\begin{array}{ll}
0 & \ \ \ \ \ \ \textnormal{if $|e\rangle\langle
e|\in\{|e_i\rangle\langle e_i|:1\le i\le r\}$}
\\
-\lim \frac{1}{t}\log a_{i,t} & \ \ \ \ \ \ \textnormal{if
$|e\rangle\langle e|=|e_i\rangle\langle e_i|$,  $r+1\le i\le N$}
 \\
+\infty & \ \ \ \ \ \ elsewhere.
\end{array}
\right.
\end{equation}
In particular, for each $i\in\{1,...,N\}$ and for each open set
$G\subset P$ containing $\omega_{|e_i\rangle\langle e_i|}$ such that
$\overline{G}\cap\{\omega_{|e_j\rangle\langle e_j|}:1\le j\le N,j\neq
i\}=\emptyset$,  $\lim\frac{1}{t}\log\mu_t(G)$ exists and satisfies
\[\lim\frac{1}{t}\log\mu_t(G)=\lim\frac{1}{t}\log a_{i,t}.\]
\item[(b)] Conversely, if $(\omega_{\rho_t})$ is
represented by a net of
 orthogonal measures $(\mu_t)$ satisfying a large deviation principle
 with rate function (\ref{LDP-converging-eq0}) (where $\rho_\infty=\sum_{i=1}^N
a_i|e_i\rangle\langle e_i|$), then   $(P_i)$ holds for all
  $i$ where $J(\omega_{|e_i\rangle\langle e_i|})<+\infty$.
\end{itemize}
\end{prop}

  \begin{pf}
 Assume that  $(P_i)$ holds for all $i\in\{1,...,N\}$.
  The convergence of states implies
 $\lim||\rho_t-\rho_\infty||=0$ so that
$\lim
 a_{i,t}=a_i$ for all $i\in\{1,...,N\}$, and in particular
$\lim a_{i,t}^{1/t}=1$ when $1\le i\le r$. Assume that
 $\lim a_{i,t}^{1/t}$ exists for all
$i\in\{r+1,...,N\}$.
   All
the hypotheses of Proposition \ref{prop-conditions} hold, and the
large deviations  follow for  $(\mu_t)$, with rate function given by
(\ref{prop-conditions-eq0}). Let $\omega_{|e\rangle\langle e|}\in
P$. For each $h\in\mathcal{A}$ with $h(\omega_{|e\rangle\langle
e|})=0$ there exist $\omega'\in P$ and a finite set $K\subset M_N$
such that $h=\inf_{x\in K}\{-|\hat{x}-\hat{x}(\omega')|\}$ and
$\omega_{|e\rangle\langle e|}(x)=\omega'(x)$ for all $x\in K$. We
put $h_{K,\omega'}=h$, and $c_i=\lim \frac{1}{t}\log a_{i,t}$ for
all $i\in\{1,...,N\}$. By (\ref{prop-conditions-eq1}) we have
\begin{equation}\label{LDP-converging-eq1}
\Lambda(h_{K,\omega'})=\sup_{1\le i\le N}(c_i-\sup_{x\in
K}\lim|\langle e_{i,t},x e_{i,t}\rangle-\omega'(x)|)=\sup_{1\le i\le
N}(c_i-\sup_{x\in K}|\langle e_i,x e_i\rangle-\langle e,x
e\rangle|),\end{equation} so that $\Lambda(h)=0$ for all
$h\in\mathcal{A}$ when $|e\rangle\langle e|\in\{|e_i\rangle\langle
e_i|:1\le i\le r\}$ (since in this case $c_i=0$), hence
$J(\omega_{|e\rangle\langle e|})=0$ by (\ref{prop-conditions-eq0}).
Assume now that $|e\rangle\langle e|\not\in\{|e_i\rangle\langle
e_i|:1\le i\le N\}$. Let $x\in M_N$ such that $\langle e_i,x
e_i\rangle\neq\langle e,x e\rangle$ for all $i\in\{1,...,N\}$, and
put $\delta=\inf_{1\le i\le N}|\langle e_i,x e_i\rangle-\langle e,x
e\rangle|$. For all $M>0$ there exists $r_M>0$ such that
$\inf|\langle e_i,r_M x e_i\rangle-\langle e,r_M x e\rangle|>M$, and
since by (\ref{LDP-converging-eq1})
\[-\Lambda(h_{\{r_M x\},\omega'})\ge
\inf_{1\le i\le N}|\langle e_i,r_M x e_i\rangle-\langle e,r_M x
e\rangle|,\] we get by letting $M\rightarrow+\infty$,
\[+\infty=\sup_M\{-\Lambda(h_{\{r_M x\},\omega'})\}\le
 \sup_{h\in\mathcal{A},h(\omega_{|e\rangle\langle
 e|})=0}\{-\Lambda(h)\},\] that is  $J(\omega_{|e\rangle\langle
 e|})=+\infty$. Assume now that $|e\rangle\langle
e|=|e_i\rangle\langle e_i|$ for some $i\in\{r+1,...,N\}$.
 By the extended version of Varadhan' theorem for
 $[-\infty,+\infty[$-valued bounded above functions (see Corollary 3.2 of \cite{com1}),
large deviations imply that $\lim\mu_{t}^{1/t}({\widehat{x}}^t)$
exists for all positive $x\in M_N$,  and satisfies
$\lim\mu_{t}^{1/t}({\widehat{x}}^t)=\sup_{\omega\in
P}\widehat{x}(\omega)e^{-J(\omega)}$. Taking $x=|e_i\rangle\langle
e_i|$ yields
\[
\lim\mu_{t}^{1/t}(\widehat{|e_i\rangle\langle e_i|}^t)=
\sup_{u\in\mathcal{H},\|u\|=1}|\langle
u,e_i\rangle|^2e^{-J(\omega_{|u\rangle\langle u|})},
\]
and by the preceding cases we see that  the  only possible non-zero
value of the map $|u\rangle\langle u|\mapsto|\langle
u,e_i\rangle|^2e^{-J(\omega_{|u\rangle\langle u|})}$ is obtained at
the point $|e_i\rangle\langle e_i|$, so that
\[\lim\mu_{t}^{1/t}(\widehat{|e_i\rangle\langle
e_i|}^t)=e^{-J(\omega_{|e_i\rangle\langle e_i|})}.\] Since
$\lim\langle e_i,e_{j,t}\rangle=0$ for all $j\neq i$ and
$\lim\langle e_i,e_{i,t}\rangle=1$, it follows that
\[
\lim\mu_{t}^{1/t}(\widehat{|e_i\rangle\langle e_i|}^t)= \max_{1\le
j\le N}\{\lim a_{j,t}^{1/t}|\langle e_i,e_{j,t}\rangle|^2\}=\lim
a_{i,t}^{1/t}\] hence $J(\omega_{|e_i\rangle\langle
e_i|})=-\lim\frac{1}{t}\log a_{i,t}$. We have proved the "if" part
of the first assertion of $(a)$, and the second assertion. If
$(\mu_t)$ satisfies a large deviation principle
 with powers
$(1/t)$, then $\lim\mu_{t}^{1/t}(\widehat{|e_i\rangle\langle
e_i|}^t)$ exists and
\begin{equation}\label{LDP-converging-eq4}
\lim\mu_{t}^{1/t}(\widehat{|e_i\rangle\langle e_i|}^t)= \max_{1\le
j\le N}\{\limsup a_{j,t}^{1/t}|\langle e_i,e_{j,t}\rangle|^2\}=\lim
a_{i,t}^{1/t}
\end{equation}
 for all $i\in\{r+1,...,N\}$; this proves the "only" part of the first assertion of $(a)$.
 The last assertion follows from  (\ref{subsection-LDP-eq1}) since
  (\ref{LDP-converging-eq0}) implies that
  any open set $G\subset P$ containing $|e_i\rangle\langle e_i|$
  with
$\overline{G}\cap\{|e_j\rangle\langle e_j|:1\le j\le N,j\neq
i\}=\emptyset$, is a $J$-continuity set. The proof of $(a)$ is
complete.

Assume  that the hypotheses of $(b)$ hold. The extended version of
Varadhan's theorem together with (\ref{LDP-converging-eq0}) yield
for each $i\in\{1,...,N\}$,
\begin{equation}\label{LDP-converging-eq6}
\lim\mu_{t}^{1/t}(\widehat{|e_i\rangle\langle e_i|}^t)=\max_{1\le
j\le N}\{\limsup a_{j,t}^{1/t}|\langle
e_i,e_{j,t}\rangle|^2\}
\end{equation}
\[=\sup_{u\in\mathcal{H},\|u\|=1}|\langle
u,e_i\rangle|^2e^{-J(\omega_{|u\rangle\langle u|})}=\lim
a_{i,t}^{1/t}.\] Let $s$ be the greatest integer less than $N$ such that $\lim
a_{i,t}^{1/t}>0$ for all $i\le s$, and note that $s\ge r\ge 1$.
 By (\ref{LDP-converging-eq6}) we have  $\lim a_{1,t}^{1/t}=\lim a_{j,t}^{1/t}|\langle
e_1,e_{j,t}\rangle|^2$ for some $j\in\{1,...,N\}$. If $j=1$, then
clearly  $\lim|\langle e_1,e_{1,t}\rangle|=1$ and the conclusion
holds. If $s=1$ the proof is complete so let us assume that $s>1$.
Assume that $j>1$. Since $\lim a_{1,t}^{1/t}\ge\lim a_{j,t}^{1/t}$
we have necessarily $\lim a_{1,t}^{1/t}=\lim a_{j,t}^{1/t}$ and
$\lim|\langle e_1,e_{j,t}\rangle|=1$. With the new labeling of the
eigenvalues of $\rho_t$ obtained interchanging $(1,t)$ and $(j,t)$,
the conclusion holds for the first eigenvector. Suppose that the
result holds for  all $i-1<s$, and assume that
\[\lim a_{i,t}^{1/t}=\lim a_{j,t}^{1/t}|\langle
e_{i},e_{j,t}\rangle|^2\] for some $j\in\{1,...,N\}$.
 Since $\lim|e_{j,t}\rangle\langle
e_{j,t}|=|e_j\rangle\langle e_j|$ for all $j\le i-1$ and $\lim
a_{i,t}^{1/t}>0$, it follows that $j\ge i$, and the proof is the
same as for the case $i=1$. We conclude with a finite recurrence
(note that at each step   the new labeling preserves the
inequalities $\lim a_{i,t}^{1/t}\ge\lim a_{j,t}^{1/t}$ when $j\ge
i$).
\end{pf}

\begin{prop}\label{LDP-implies-conv}
Let  $(\omega_t)_{t\ge 0}$ be a net of states on $M_N$. If
$(\omega_t)$ is represented by a  net of orthogonal measures which
satisfies a large deviation principle with a rate function vanishing
at a unique point  $\omega_{|e\rangle\langle e|}$, then
$\lim\omega_t=\omega_{|e\rangle\langle e|}$.
\end{prop}

\begin{pf}
 Let $\mu_t=\sum_{1\le i\le N}
 a_{i,t}\delta_{\omega_{|e_{i,t}\rangle\langle e_{i,t}|}}$ be the
 orthogonal measure representing $\omega_t$ as
above,
  let $\varepsilon>0$, let
$x\in M_N$ with $||x||=1$,  and let
 $G$ be the  open neighborhood of $\omega_{|e\rangle\langle e|}$ defined by
 $\{\omega\in P:|\omega(x)-\omega_{|e\rangle\langle
e|}(x)|<\varepsilon\}$.
 Then,
 \begin{equation}\label{LDP-implies-conv-eq-20}
 \mu_t(\hat{x})=\sum_{1\le i\le N, \omega_{|e_{i,t}\rangle\langle e_{i,t}|}\in G}
 a_{i,t}\omega_{|e_{i,t}\rangle\langle e_{i,t}|}(x)+
\sum_{1\le i\le N, \omega_{|e_{i,t}\rangle\langle e_{i,t}|}\not\in
G}
 a_{i,t}\omega_{|e_{i,t}\rangle\langle e_{i,t}|}(x)
\end{equation}
and note that
\begin{equation}\label{LDP-implies-conv-eq-21}
\sum_{1\le i\le N, \omega_{|e_{i,t}\rangle\langle e_{i,t}|}\not\in
G}
 a_{i,t}\omega_{|e_{i,t}\rangle\langle e_{i,t}|}(x)
\le
 \mu_t(P\verb'\'G).
 \end{equation}
We have
\[\omega_{|e\rangle\langle e|}(x)
\sum_{1\le i\le N, \omega_{|e_{i,t}\rangle\langle e_{i,t}|}\in G}
 a_{i,t}-\varepsilon \sum_{1\le i\le N, \omega_{|e_{i,t}\rangle\langle e_{i,t}|}\in G}
 a_{i,t}\le
\]
\[\sum_{1\le i\le N, \omega_{|e_{i,t}\rangle\langle e_{i,t}|}\in G}
 a_{i,t}\omega_{|e_{i,t}\rangle\langle e_{i,t}|}(x)
\le\sum_{1\le i\le N, \omega_{|e_{i,t}\rangle\langle e_{i,t}|}\in G}
 a_{i,t}(\omega_{|e\rangle\langle e|}(x)+\varepsilon)\]
 \[\le\omega_{|e\rangle\langle e|}(x)+\varepsilon.\]
  Let $\varepsilon\rightarrow 0$ and get
\begin{equation}\label{LDP-implies-conv-eq-22}
\omega_{|e\rangle\langle e|}(x)\sum_{1\le i\le N,
\omega_{|e_{i,t}\rangle\langle e_{i,t}|}\in G}
 a_{i,t}\le\sum_{1\le i\le N, \omega_{|e_{i,t}\rangle\langle e_{i,t}|}\in G}
 a_{i,t}\omega_{|e_{i,t}\rangle\langle e_{i,t}|}(x)\le\omega_{|e\rangle\langle
e|}(x)
 \end{equation}
The large deviations and the  hypothesis on the rate function imply
that
  $\lim\mu_t(P\verb'\'G)=0$ (exponentially fast) hence
   \[\lim_t\sum_{1\le i\le N, \omega_{|e_{i,t}\rangle\langle e_{i,t}|}\in G}
 a_{i,t}=\lim_t\mu_t(G)=1,\] and (\ref{LDP-implies-conv-eq-22})
 yields
 \begin{equation}\label{LDP-implies-conv-eq-23}
 \lim_t\sum_{1\le i\le N, \omega_{|e_{i,t}\rangle\langle e_{i,t}|}\in G}
 a_{i,t}\omega_{|e_{i,t}\rangle\langle e_{i,t}|}(x)=\omega_{|e\rangle\langle
e|}(x).
 \end{equation}
Then (\ref{LDP-implies-conv-eq-20}), (\ref{LDP-implies-conv-eq-21}),
(\ref{LDP-implies-conv-eq-23}) give
$\lim\mu_t(\hat{x})=\omega_{|e\rangle\langle e|}(x)$, which proves
the proposition by Lemma \ref{conv-meas-conv-states}.
\end{pf}

\section{The case of states
arising from quantum Markov semigroups on
$M_2$}\label{section-Markov}

In this section, we first prove a  general property of quantum
Markov semigroups on $M_2$ having an absorbing state (Theorem
\ref{geom-prop}). Looking the state space of $M_2$ as the unit ball
in $\mathbb{R}^3$, it says that when the absorbing state
$\omega_\infty$ is on the sphere (i.e., pure),
$\mathcal{T}_{*t}(\omega)$ approaches  $\omega_\infty$ remaining
inside the open unit ball, for all states $\omega\neq\omega_\infty$. This property is
crucial for the proof of the large deviations result (Theorem
\ref{finite-case}).

\begin{thm}\label{geom-prop}
Let $(\mathcal{T}_t)$ be a quantum Markov semigroup acting on $M_2$,
and having   an absorbing state $\omega_\infty$. Then, for each
state $\omega\neq\omega_\infty$, $\mathcal{T}_{*t}(\omega)$ is
faithful for all $t$ large enough.
\end{thm}

\begin{pf}
 Clearly, the conclusion holds when  $\omega_\infty$ is faithful, so that
let us assume that $\omega_\infty$ is pure given by some projection
$|e_1\rangle\langle e_1|$, and let $e_2$ be a unit vector orthogonal
to $e_1$. First assume that $\omega$ is a pure state given  by a
unit vector $e=\alpha_1 e_1+\alpha_2 e_2$ with $\alpha_2\neq 0$, and
suppose the conclusion does not hold.  There exists a sequence
$(t_n)$ such that $\tilde\mathcal{T}_{*t_n}(|e\rangle\langle e|)$ is
a rank-one projection for all $n\in\mathbb{N}$. Let
$(\mathcal{S}_t),\mathcal{J},y, z_1, z_2$ as in Theorem
\ref{Davies-result}, and note that  $y e_1=y^* e_1=0$ implies that
$y$ is normal and diagonalizes in the basis $\{e_1,e_2\}$, hence $y
e_2=\gamma e_2$ for some $\gamma\in\mathbb{C}$. Put
 $-\eta=\gamma+\overline{\gamma}$, and note that $-\eta<0$ since $-\eta$ is the least
eigenvalue of $y+y^*$. For each $n\in\mathbb{N}$, let $u_n$ be a
unit vector such that $\langle
u_n,\tilde\mathcal{T}_{*t_n}(|e\rangle\langle e|)u_n\rangle=0$, and
get by  (\ref{Davies-result-eq3}) and (\ref{Davies-result-eq4})
\[\sum_{i=1}^2\int_0^{t_n}\langle u_n,\tilde\mathcal{T}_{*t_n-s}
(|z_ie^{sy}e\rangle\langle z_ie^{sy}e|)u_n\rangle ds=0,\] and so for
$i\in\{1,2\}$ and for each $s\in[0,t_n]$
\[
\langle u_n,\tilde\mathcal{T}_{*t_n-s} (|z_ie^{sy}e\rangle\langle
z_ie^{sy}e|)u_n\rangle=0.
\]
 Since
\[|e^{sy}e\rangle\langle
e^{sy}e|=|\alpha_1|^2|e_1\rangle\langle
e_1|+|\alpha_2|^2e^{-s\eta}|e_2\rangle\langle e_2|+
\alpha_1\overline{\alpha_2}e^{s\overline{\gamma}}|e_1\rangle\langle
e_2|+ \alpha_2\overline{\alpha_1}e^{s\gamma}|e_2\rangle\langle
e_1|\] we get for $i\in\{1,2\}$
\[|z_ie^{sy}e\rangle\langle z_ie^{sy}e|=|\alpha_2|^2e^{-s\eta}|z_ie_2\rangle\langle
 z_ie_2|\]hence
\begin{equation}\label{geom-prop-eq3}
\forall n\in\mathbb{N}, \forall s\in[0,t_n]\ \ \ \ \ \ \langle
u_n,\tilde\mathcal{T}_{*s}(|z_ie_2\rangle\langle z_i e_2|)
u_n\rangle=0.
\end{equation} Put $z_ie_2=\alpha_{1,i} e_1+\alpha_{2,i}
e_2$ for $i\in\{1,2\}$ and get
\[
\tilde\mathcal{T}_{*s}(|z_ie_2\rangle\langle
z_ie_2|)=|\alpha_{1,i}|^2|e_1\rangle\langle
e_1|+|\alpha_{2,i}|^2\tilde\mathcal{T}_{*s}(|e_2\rangle\langle
e_2|)+\overline{\alpha_{2,i}}\alpha_{1,i}\tilde\mathcal{T}_{*s}(|e_1\rangle\langle
e_2|)\]
\begin{equation}\label{geom-prop-eq4}
+\overline{\alpha_{1,i}}\alpha_{2,i}\tilde\mathcal{T}_{*s}(|e_2\rangle\langle
e_1|).\end{equation}
 By (\ref{Davies-result-eq4}) we have
\[
\tilde\mathcal{T}_{*s}(|e_2\rangle\langle
e_2|)=e^{-s\eta}|e_2\rangle\langle
e_2|+\sum_{i=1}^2\int_0^s\tilde\mathcal{T}_{*s-r}(z_i|e^{r\gamma
}e_2\rangle\langle e^{r\gamma} e_2|z_i^*)dr
\]
\begin{equation}\label{geom-prop-eq5}
=e^{-s\eta}|e_2\rangle\langle e_2|+ \sum_{i=1}^2\int_0^s
e^{-r\eta}\tilde\mathcal{T}_{*s-r}(|z_ie_2\rangle\langle z_ie_2|)dr.
\end{equation}
 Then,
 \[
 \tilde\mathcal{T}_{*s}(|e_1\rangle\langle
e_2|)=\mathcal{S}_s(|e_1\rangle\langle
e_2|)+\sum_{i=1}^2\int_0^s\tilde\mathcal{T}_{*s-r}(z_i|e^{ry}e_1\rangle\langle
e^{ry} e_2|z_i^*)dr=\]
\begin{equation}\label{geom-prop-eq5.1}
\mathcal{S}_s(|e_1\rangle\langle
e_2|)+\sum_{i=1}^2\int_0^s\tilde\mathcal{T}_{*s-r}(|z_ie_1\rangle\langle
z_ie^{r\gamma} e_2|)=\mathcal{S}_s(|e_1\rangle\langle e_2|)=
e^{s\overline{\gamma}}|e_1\rangle\langle e_2|.
\end{equation}
 In the same way
we get
\begin{equation}\label{geom-prop-eq5.2}
 \tilde\mathcal{T}_{*s}(|e_2\rangle\langle
e_1|)=e^{s\gamma}|e_2\rangle\langle e_1|.
\end{equation}
Then (\ref{geom-prop-eq4}), (\ref{geom-prop-eq5}),
(\ref{geom-prop-eq5.1}), (\ref{geom-prop-eq5.2})  give for each
 $n\in\mathbb{N}$,
\[
\langle u_n,\tilde\mathcal{T}_{*s}(|z_ie_2\rangle\langle
z_ie_2|)u_n\rangle=|\alpha_{1,i}|^2|\langle
e_1,u_n\rangle|^2++2\mathcal{R}\textnormal{e}
(\overline{\alpha_{1,i}}\alpha_{2,i}e^{s\gamma}\langle
e_1,u_n\rangle\langle u_n,e_2\rangle)\]
\[+|\alpha_{2,i}|^2\langle
u_n,\tilde\mathcal{T}_{*s}(|e_2\rangle\langle e_2|)u_n\rangle
\]
\begin{equation}\label{geom-prop-eq6}
=|\alpha_{1,i}|^2|\langle
e_1,u_n\rangle|^2+2\mathcal{R}\textnormal{e}
(\overline{\alpha_{1,i}}\alpha_{2,i}e^{s\gamma}\langle
e_1,u_n\rangle\langle
u_n,e_2\rangle)+|\alpha_{2,i}|^2e^{-s\eta}(|\langle
e_2,u_n\rangle|^2\]
\[+|\alpha_{2,i}|^2\sum_{j=1}^2\int_0^s e^{-r\eta}\langle u_n,
\tilde\mathcal{T}_{*s-r}(|z_je_2\rangle\langle z_je_2|)u_n\rangle
dr.
\end{equation}
By (\ref{geom-prop-eq3}) and  (\ref{geom-prop-eq6}) we obtain for
each $n\in\mathbb{N}$ and each $s\in[0,t_n]$
\begin{equation}\label{geom-prop-eq7}
|\alpha_{1,i}|^2|\langle
e_1,u_n\rangle|^2+2\mathcal{R}\textnormal{e}
(\overline{\alpha_{1,i}}\alpha_{2,i}e^{s\gamma}\langle
e_1,u_n\rangle\langle
u_n,e_2\rangle)+|\alpha_{2,i}|^2e^{-s\eta}|\langle
e_2,u_n\rangle|^2=0.
\end{equation}
Taking the limit when $n\rightarrow+\infty$  in
(\ref{geom-prop-eq7}) with $s=t_n$ yields
\begin{equation}\label{geom-prop-eq8}
\lim_n|\alpha_{1,i}|^2|\langle e_1,u_n\rangle|^2=0,
\end{equation}and (\ref{geom-prop-eq3})
with $s=0$ gives
\begin{equation}\label{geom-prop-eq9}
\forall n\in\mathbb{N},\ \ \ \ \ {\alpha_{1,i}}\langle
e_1,u_n\rangle+{\alpha_{2,i}}\langle e_2,u_n\rangle=0.
\end{equation}
 First assume $\alpha_{2,i}=0$. If
$\alpha_{1,i}\neq 0$, then $|u_n\rangle\langle
u_n|=|e_2\rangle\langle e_2|$ for all $n\in\mathbb{N}$ by
(\ref{geom-prop-eq9}), and (\ref{Davies-result-eq3}) implies
\[\forall n\in\mathbb{N},\ \ \ \ \ \langle e_2, S_{t_n}(|e\rangle\langle e|)e_2\rangle
=\langle e^{t_n\overline{\gamma}}e_2, |e\rangle\langle
e|e^{t_n\overline{\gamma}}e_2\rangle= |\alpha_2|^2e^{-t_n\eta}=0\]
which gives the contradiction since $\alpha_2\neq 0$. It follows
that $\alpha_{1,i}=0$,
  and so
$z_i=0$. Assume now $\alpha_{2,i}\neq 0$. It is easy to see that
(\ref{geom-prop-eq8}) and (\ref{geom-prop-eq9}) imply $\lim\langle
e_2,u_n\rangle=0$ and  $\alpha_{1,i}= 0$,  hence
   $|u_n\rangle\langle
u_n|=|e_1\rangle\langle e_1|$ for all $n\in\mathbb{N}$, which gives
the contradiction since $\lim\langle
e_1,\tilde\mathcal{T}_{*t_n}(|e\rangle\langle e|)e_1\rangle=1$. We
obtain finally that (\ref{geom-prop-eq8}) and (\ref{geom-prop-eq9})
imply $z_1=z_2=0$, that is $\tilde\mathcal{T}_{*t}=S_t$ for all
$t\ge 0$. Since
$\min\sigma(\tilde\mathcal{T}_{*t_n}(|e\rangle\langle e|))=0$ and
$\textnormal{tr}(\tilde\mathcal{T}_{*t_n}(|e\rangle\langle e|))=1$,
we have
\[1=||S_{t_n}(|e\rangle\langle e|)||=\sup_{u\in\mathcal{H},||u||=1}
\{|\alpha_1|^2|\langle
u,e_1\rangle|^2+|\alpha_2|^2e^{-t_n\eta}|\langle u, e_2\rangle|^2\]
\[+2\mathcal{R}\textnormal{e}(\overline{\alpha_1}\alpha_2e^{t_n\gamma}\langle
e_1,u\rangle\langle u, e_2\rangle)\},\] and letting
$t_n\rightarrow+\infty$ it follows that
\[1=\sup_{u\in\mathcal{H},||u||=1}
\{|\alpha_1|^2|\langle u,e_1\rangle|^2\},\] which implies
$|\alpha_1|=1$ and the contradiction since $\alpha_2\neq 0$. The
theorem is proved when $\omega$ is pure. Assume now that $\omega$ is
given by some strictly positive operator $\rho$, i.e. $cI\le \rho\le
I$ for some $c>0$. Since $\mathcal{S}_t(cI)=ce^{t(y+y^*)}$,
(\ref{Davies-result-eq3}) gives
\[
ce^{-t\eta}=\min\sigma(\mathcal{S}_t(cI))\le\min\sigma(\tilde\mathcal{T}_{*t}(\rho)),
\]
which shows that $\mathcal{T}_{*t}(\omega)$ is faithful for all
$t\ge 0$.
\end{pf}

The following theorem is our large deviation result. It shows that
when $(\mathcal{T}_t)$ has an absorbing pure state $\omega_\infty$,
and for any initial state $\omega_\rho\neq\omega_\infty$, the net
$(\mathcal{T}_{*t}(\omega_\rho))$ converges exponentially fast (in
the large deviation sense expressed by (\ref{finite-case-eq1}))
toward $\omega_\infty$ with rate $\eta-a$, where $\eta,a$ are
parameters given by the generator. Note that (\ref{finite-case-eq1})
implies $a_{2,t,\rho}>0$ for all $t$ large enough, so that Theorem
\ref{finite-case} contains Theorem \ref{geom-prop}.

\begin{thm}\label{finite-case}
Let $(\mathcal{T}_t)$ be a quantum Markov semigroup on $M_2$, and
having an absorbing pure state $\omega_{|e_1\rangle\langle e_1|}$.
 Let $y,z_1,z_2$ be the parameters of the generator
$\tilde\mathcal{L}_*$ as in Theorem \ref{Davies-result}, and let
$e_2$ be  a unit vector orthogonal to $e_1$.
  Then for each state
$\omega_\rho\neq\omega_{|e_1\rangle\langle e_1|}$, the net
$(\mu_{t,\rho})$ of orthogonal measures representing
$(\mathcal{T}_{*t}(\omega_\rho))$ satisfies a large deviation
principle with powers $(1/t)$ and rate function
\begin{displaymath}
 J(\omega_{|e\rangle\langle
e|})=\left\{
\begin{array}{ll}
0 & \ \ \ \ \ \ \textnormal{if $|e\rangle\langle
e|=|e_1\rangle\langle e_1|$}
\\
\eta-a & \ \ \ \ \ \ \textnormal{if $|e\rangle\langle
e|=|e_2\rangle\langle e_2|$}
\\
+\infty & \ \ \ \ \ \ elsewhere,
\end{array}
\right.
\end{displaymath}
where $a=|\langle z_1e_2,e_2\rangle|^2+|\langle
z_2e_2,e_2\rangle|^2$ and $\eta$ is  the greatest eigenvalue of
$-(y+y^*)$; moreover, $\eta-a>0$. In particular,  the rate function
does not depend on the initial state $\omega_\rho$,  and for each
open sets $G\subset P$ containing $|e_2\rangle\langle e_2|$ such
that $|e_1\rangle\langle e_1|\not\in\overline{G}$,
$\lim\frac{1}{t}\log\mu_{t,\rho}(G)$ exists and satisfies
\begin{equation}\label{finite-case-eq1}
\lim\frac{1}{t}\log\mu_{t,\rho}(G)=\lim\frac{1}{t}\log
a_{2,t,\rho}=a-\eta,
\end{equation} where $a_{2,t,\rho}$ is the
least eigenvalue of $\tilde\mathcal{T}_{*t}(\rho)$.
\end{thm}

\begin{pf}
 Let $\omega_\rho\neq\omega_{|e_1\rangle\langle e_1|}$  be a state,  and  let $\tilde\mathcal{T}_{*t}(\rho)$ have the diagonal form
 $\tilde\mathcal{T}_{*t}(\rho)=\sum_{i=1}^2 a_{i,t,\rho}|e_{i,t,\rho}\rangle\langle
e_{i,t,\rho}|$  as in Lemma \ref{prop-eigen}. By Proposition
\ref{LDP-converging} we only have to check that $\lim
a_{2,t,\rho}^{1/t}$ exists and equals $e^{a-\eta}$. Since
\begin{equation}\label{finite-case-eq3}
\tilde\mathcal{T}_{*t}(I)=\tilde\mathcal{T}_{*t}(|e_1\rangle\langle
e_1|)+\tilde\mathcal{T}_{*t}(|e_2\rangle\langle
e_2|)=|e_1\rangle\langle
e_1|+\tilde\mathcal{T}_{*t}(|e_2\rangle\langle e_2|),
\end{equation}
 by (\ref{geom-prop-eq5}) we have
\begin{equation}\label{finite-case-eq7}
\min\sigma(\tilde\mathcal{T}_{*t}(I))\le \langle
e_2,\tilde\mathcal{T}_{*t}(I)e_2\rangle=\langle
e_2,\tilde\mathcal{T}_{*t}(|e_2\rangle\langle
e_2|)e_2\rangle\end{equation}
\[=e^{-t\eta}+\sum_{i=1}^2\int_0^t
e^{-s\eta}\langle e_2,\tilde\mathcal{T}_{*t-s}(|z_ie_2\rangle\langle
z_ie_2|)e_2\rangle\] and by using the first equality of
(\ref{geom-prop-eq6}) (with $e_2$ in place of $u_n$ and $t-s$ in
place of $s$), the last above equality becomes
\[
\langle e_2,\tilde\mathcal{T}_{*t}(|e_2\rangle\langle
e_2|)e_2\rangle=e^{-t\eta}+a\int_0^t e^{-s\eta}\langle
e_2,\tilde\mathcal{T}_{*t-s}(|e_2\rangle\langle e_2|)e_2\rangle
\]
with $a=|\alpha_{2,1}|^2+|\alpha_{2,2}|^2$. By putting $u(t)=\langle
e_2,\tilde\mathcal{T}_{*t}(|e_2\rangle\langle e_2|)e_2\rangle$ for
all $t\ge 0$, and  applying the Laplace transform, it is easy to see
that the equation
\[
u(t)=e^{-t\eta}+a\int_0^t  e^{-s\eta} u(t-s)ds
\]
has the  unique solution
\begin{equation}\label{finite-case-eq10}
\forall t\ge 0,\ \ \ \ \ \ u(t)=e^{t(a-\eta)},
\end{equation} with $a-\eta<0$ (since $\lim\langle
e_2,\tilde\mathcal{T}_{*t}(|e_2\rangle\langle e_2|)e_2\rangle=0$).
It follows from (\ref{finite-case-eq7}) that
\begin{equation}\label{finite-case-eq10.00}
\limsup\min\sigma(\tilde\mathcal{T}_{*t}(I))^{1/t}\le \lim\langle
e_2,\tilde\mathcal{T}_{*t}(|e_2\rangle\langle
e_2|)e_2\rangle^{1/t}=e^{a-\eta}<1.
\end{equation}
 For each $t\ge
0$ let $u_t$ be a unit vector such that
\[\min\sigma(\tilde\mathcal{T}_{*t}(I))=\langle
u_t,\tilde\mathcal{T}_{*t}(I)u_t\rangle.\] By
(\ref{finite-case-eq3}) we have
\[\min\sigma(\tilde\mathcal{T}_{*t}(I))=|\langle u_t, e_1\rangle|^2
+\langle u_t,\tilde\mathcal{T}_{*t}(|e_2\rangle\langle
e_2|)u_t\rangle.\]
 Let $(u_{t_j})$ be a subnet of $(u_t)$ such
that
\[\liminf\min\sigma(\tilde\mathcal{T}_{*t}(I))^{1/t}=
\lim\min\sigma(\tilde\mathcal{T}_{*t_j}(I))^{1/t_j}\] and get
\begin{equation}\label{finite-case-eq10.0}
\liminf\min\sigma(\tilde\mathcal{T}_{*t}(I))^{1/t}=\limsup|\langle
u_{t_j}, e_1\rangle|^{2/t_j} \vee\limsup\langle
u_{t_j},\tilde\mathcal{T}_{*t_j}(|e_2\rangle\langle
e_2|)u_{t_j}\rangle^{1/t_j}.
\end{equation}
Let $(u_{t_{k}})$ be a subnet of $(u_{t_{j}})$. Then
$|u_{t_k}\rangle\langle u_{t_k}|$ has a subnet
$|u_{t_l}\rangle\langle u_{t_l}|$ converging to some  projection
$|u\rangle\langle u|$. If $|u\rangle\langle u|\neq
|e_2\rangle\langle e_2|$, then
$\liminf\min\sigma(\tilde\mathcal{T}_{*t}(I))^{1/t}=1$,  which
contradicts (\ref{finite-case-eq10.00}).
 Therefore,
$|u\rangle\langle u|=|e_2\rangle\langle e_2|$ and since the subnet
$|u_{t_k}\rangle\langle u_{t_k}|$  is arbitrary, $|u_t\rangle\langle
u_t|$ converges to $|e_2\rangle\langle e_2|$.
 Put $u_t=b_{1,t}e_1+b_{2,t}e_2$ for
all $t\ge 0$, and get
\begin{equation}\label{finite-case-eq10.1}
\langle u_{t_j},\tilde\mathcal{T}_{*t_j}(|e_2\rangle\langle
e_2|)u_{t_j}\rangle=|b_{1,t_j}|^2\langle
e_1,\tilde\mathcal{T}_{*t_j}(|e_2\rangle\langle
e_2|)e_1\rangle+|b_{2,t_j}|^2\langle
e_2,\tilde\mathcal{T}_{*t_j}(|e_2\rangle\langle e_2|)e_2\rangle
\end{equation}
\[+\overline{b_{1,t_j}}b_{2,t_j}\langle
e_1,\tilde\mathcal{T}_{*t_j}(|e_2\rangle\langle
e_2|)e_2\rangle+\overline{b_{2,t_j}}b_{1,t_j}\langle
e_2,\tilde\mathcal{T}_{*t_j}(|e_2\rangle\langle e_2|)e_1\rangle.\]
Then  (\ref{geom-prop-eq5}) combined with (\ref{geom-prop-eq4}),
(\ref{geom-prop-eq5.1}), (\ref{geom-prop-eq5.2}) yield
\[
\forall t\ge 0,\ \ \ \ \ \langle
e_1,\tilde\mathcal{T}_{*t}(|e_2\rangle\langle
e_2|)e_2\rangle=a\int_0^t e^{-s\eta}\langle
e_1,\tilde\mathcal{T}_{*t-s}(|e_2\rangle\langle e_2|)e_2\rangle ds,
\]
and  by an application of Laplace transform we get as unique
solution
\begin{equation}\label{finite-case-eq11}
\langle e_1,\tilde\mathcal{T}_{*t}(|e_2\rangle\langle
e_2|)e_2\rangle=0\ \ \ \ \ \textnormal{for all $t\ge 0$}.
\end{equation}
Similar calculations yield
\begin{equation}\label{finite-case-eq11.2}
\langle e_2,\tilde\mathcal{T}_{*t}(|e_2\rangle\langle
e_2|)e_1\rangle=0\ \ \ \ \ \textnormal{for all $t\ge 0$}.
\end{equation}
 It follows from (\ref{finite-case-eq10}) and
(\ref{finite-case-eq10.1}) that
\[\langle u_{t_j},\tilde\mathcal{T}_{*t_j}(|e_2\rangle\langle
e_2|)u_{t_j}\rangle\ge |b_{2,t_j}|^2e^{t_j(a-\eta)}.\] Since $\lim
|u_{t_j}\rangle\langle u_{t_j}|=|e_2\rangle\langle e_2|$, we have
$\lim|b_{2,t_j}|=1$ and by (\ref{finite-case-eq10.0})
\begin{equation}\label{finite-case-eq12}
\liminf\min\sigma(\tilde\mathcal{T}_{*t}(I))^{1/t}\ge\limsup\langle
u_{t_j},\tilde\mathcal{T}_{*t_j}(|e_2\rangle\langle
e_2|)u_{t_j}\rangle^{1/t_j}\ge e^{a-\eta}.
\end{equation}
Then (\ref{finite-case-eq10.00}) and (\ref{finite-case-eq12}) yield
\begin{equation}\label{finite-case-eq13}
\lim\min\sigma(\tilde\mathcal{T}_{*t}(I))^{1/t}=\lim\langle
e_2,\tilde\mathcal{T}_{*t}(|e_2\rangle\langle
e_2|)e_2\rangle^{1/t}=e^{(a-\eta)}.
\end{equation}
 If $\rho$ is strictly positive, then $cI\le \rho\le I$ and
 \[c\min(\sigma(\tilde\mathcal{T}_{*t}(I)))\le
 \min(\sigma(\tilde\mathcal{T}_{*t}(\rho)))\le
 \min(\sigma(\tilde\mathcal{T}_{*t}(I)))\]for some
 $c>0$,
 hence
\begin{equation}\label{finite-case-eq14}
\lim
a_{2,t,\rho}^{1/t}=\lim\min(\sigma(\tilde\mathcal{T}_{*t}(\rho)))^{1/t}=
\lim\min\sigma(\tilde\mathcal{T}_{*t}(I))^{1/t}=e^{a-\eta}.
\end{equation}
 Assume now that $\rho=|e\rangle\langle e|$ for some unit vector $e=\alpha_1 e_1+\alpha_2
 e_2$ with $\alpha_2\neq 0$.
Let  $(a_{2,t_j,\rho}^{1/t_j})$ be a subsequence of
$(a_{2,t,\rho}^{1/t})$, and consider the corresponding subsequence
$(\mu_{t_j,\rho})$ of $(\mu_{t,\rho})$. Note that by Theorem
\ref{geom-prop},  $a_{2,t_j,\rho}>0$ for all $j$ large enough. By a
well-known compactness result in large deviation theory (see Lemma
4.1.23 of \cite{dem}, or Corollary 5 of \cite{com2} for a general
version), $(\mu_{t_j,\rho})$ has a subsequence
$(\mu_{t_{j_k},\rho})$
 satisfying a large deviation principle, so that $\lim a_{2,t_{j_k},\rho}^{1/t_{j_k}}$
 exists by Proposition
\ref{LDP-converging}.
 Put $e^{-l}=\lim
a_{2,t_{j_k},\rho}^{1/t_{j_k}}$, and get
 for each
$k'\in\mathbb{N}$,  \[-l=\lim_k
\frac{1}{t_{j_k}-t_{j_{k'}}+t_{j_{k'}}}
\log\min\sigma(\tilde\mathcal{T}_{*t_{j_k}-t_{j_{k'}}+t_{j_{k'}}}(\rho))=\]
\[\lim_k
\frac{t_{j_k}-t_{j_{k'}}}{t_{j_k}-t_{j_{k'}}+t_{j_{k'}}}\cdot\frac{1}
{t_{j_k}-t_{j_{k'}}}\log\min\sigma
(\tilde\mathcal{T}_{*t_{j_k}-t_{j_{k'}}}(\tilde\mathcal{T}_{*t_{j_{k'}}}(\rho)))\]
\[\lim_k
\frac{1} {t_{j_k}-t_{j_{k'}}}\log\min\sigma
(\tilde\mathcal{T}_{*t_{j_k}-t_{j_{k'}}}(\tilde\mathcal{T}_{*t_{j_{k'}}}(\rho)).\]
Since  $\lim_k(t_{j_k}-t_{j_{k'}})=+\infty$, the sequence
$(\tilde\mathcal{T}_{*t_{j_k}-t_{j_{k'}}}(\tilde\mathcal{T}_{*t_{j_{k'}}}(\rho)))_{k\in\mathbb{N},
t_{j_k}\ge t_{j_k'}}$ is a subsequence of
$(\tilde\mathcal{T}_{*t}(\tilde\mathcal{T}_{*t_{j_{k'}}}(\rho))_{t>0}$.
Since $\tilde\mathcal{T}_{*t_{j_{k'}}}(\rho)$ is strictly positive
for $k'$ large enough,  (\ref{finite-case-eq14}) (with
$\tilde\mathcal{T}_{*t_{j_{k'}}}(\rho)$ in place of $\rho$) implies
$\lim
a_{2,t,\tilde\mathcal{T}_{*t_{j_{k'}}}(\rho)}^{1/t}=e^{a-\eta}$
hence
\[-l=\lim
\frac{1} {t_{j_k}-t_{j_{k'}}}\log\min\sigma
(\tilde\mathcal{T}_{*t_{j_k}-t_{j_{k'}}}(\tilde\mathcal{T}_{*t_{j_{k'}}}(\rho))=\]
\[\lim
\frac{1} {t_{j_k}-t_{j_{k'}}}\log
a_{2,t_{j_k}-t_{j_{k'}},\tilde\mathcal{T}_{*t_{j_{k'}}}(\rho)}=a-\eta.\]
It follows that  each subsequence of $(a_{2,t,\rho}^{1/t})$ has a
subsequence converging to $e^{a-\eta}$, hence $\lim
a_{2,t,\rho}^{1/t}=e^{a-\eta}$.
\end{pf}

The following result shows that  the large deviation principle as
well as the exponential rate of convergence on projections are given
by the eigenvalues of $\mathcal{J}^*(|e_1\rangle\langle e_1|)$.

\begin{cor}\label{NC-LDP}
Let $(\mathcal{T}_t)$ be a quantum Markov semigroup on $M_2$ having
an absorbing pure state $\omega_{|e_1\rangle\langle e_1|}$, let
$e_2$ be  a unit vector orthogonal to $e_1$, and let $\mathcal{J}$
be  the operator on $M_2$ appearing in the generator
(\ref{Davies-result-eq1}). Then the following conclusions hold.
\begin{itemize}
\item[(a)] $\mathcal{J}^*(|e_1\rangle\langle e_1|)=(\eta-a)|e_2\rangle\langle
e_2|$ with $\eta,a$ as in Theorem \ref{finite-case};
\item[(b)] For each state $\omega\neq\omega_{|e_1\rangle\langle e_1|}$ and each  projection $p\in
M_2\verb'\'\{0\}$ we have
\[
\lim\frac{1}{t}\log\omega(\mathcal{T}_t (p))=\left\{
\begin{array}{ll}
a-\eta & \ \ \ \ \ \ \textnormal{if $p=|e_2\rangle\langle e_2|$}
\\ \\
0 & \ \ \ \ \ \ otherwise.
\end{array}
\right.
\]
\end{itemize}
\end{cor}

\begin{pf}
Differentiating  (\ref{finite-case-eq10}), (\ref{finite-case-eq11}),
(\ref{finite-case-eq11.2}) and taking the value at $t=0$ yields
respectively $\langle e_2,\tilde{\mathcal{L}}_{*}(|e_2\rangle\langle
e_2|)e_2\rangle=a-\eta$, $\langle
e_1,\tilde{\mathcal{L}}_{*}(|e_2\rangle\langle e_2|)e_2\rangle=0$,
$\langle e_2,\tilde{\mathcal{L}}_{*}(|e_2\rangle\langle
e_2|)e_1\rangle=0$. By (\ref{finite-case-eq10}) and the preservation
of the trace we have $\langle
e_1,\mathcal{T}_{*t}(|e_2\rangle\langle
e_2|)e_1\rangle=1-e^{t(a-\eta)}$ hence  $\langle
e_1,\tilde{\mathcal{L}}_{*}(|e_2\rangle\langle
e_2|)e_1\rangle=\eta-a$. Then $(a)$ follows by noting that
\[\mathcal{J}(I)=\mathcal{J}(|e_2\rangle\langle e_2|)=\tilde{\mathcal{L}}_{*}(|e_2\rangle\langle e_2|)+
\eta|e_2\rangle\langle e_2|.\] Clearly the conclusion of $(b)$ holds
when $p$ is two dimensional, so let us assume that
$p=|e\rangle\langle e|$ for some unit vector $e$.  Since
$\omega_{|e_1\rangle\langle e_1|}$ is absorbing we have
\[\lim \omega(\mathcal{T}_t (|e\rangle\langle e|))=|\langle e,e_1
\rangle|^2\] for all states $\omega$.
 When $|e\rangle\langle e|\neq |e_2\rangle\langle e_2|$
 the above limit is strictly positive hence $\lim \omega(\mathcal{T}_t
(|e\rangle\langle e|))^{1/t}=1$ and the conclusion holds. Assume
that  $|e\rangle\langle e|= |e_2\rangle\langle e_2|$ and write
$\omega=\omega_\rho$. We have
\[\min\sigma(\tilde\mathcal{T}_{*t}(I))\le\langle
e_2,\tilde\mathcal{T}_{*t}(I) e_2\rangle=\langle
e_2,\tilde\mathcal{T}_{*t}(|e_2\rangle\langle e_2|) e_2\rangle\]
hence
\[\lim\min\sigma(\tilde\mathcal{T}_{*t}(I))^{1/t}=\lim\langle
e_2,\tilde\mathcal{T}_{*t}(I) e_2\rangle^{1/t}=e^{a-\eta}\] by
(\ref{finite-case-eq13}), and finally
\[\lim\omega_\rho(\mathcal{T}_t (|e_2\rangle\langle
e_2|))^{1/t}=\lim\langle e_2,\tilde\mathcal{T}_{*t}(\rho)
e_2\rangle^{1/t}=e^{a-\eta}\] for all $\rho$ strictly positive since
in this case $cI\le\rho\le  I$ for some $c>0$. Assume now that
$\rho=|f\rangle\langle f|$ for some unit vector $f$, and let $h$ be
a unit vector orthogonal to $f$. We have
\[\lim\langle
e_2,\tilde\mathcal{T}_{*t}(I) e_2\rangle^{1/t}=\max\{\limsup\langle
e_2,\tilde\mathcal{T}_{*t}(|f\rangle\langle f|)
e_2\rangle^{1/t},\limsup\langle
e_2,\tilde\mathcal{T}_{*t}(|h\rangle\langle h|) e_2\rangle^{1/t}\}\]
\[=e^{a-\eta}\]hence
\[
\liminf\min\sigma(\tilde\mathcal{T}_{*t}(|f\rangle\langle
f|))^{1/t}\le \liminf\langle
e_2,\tilde\mathcal{T}_{*t}(|f\rangle\langle f|) e_2\rangle^{1/t}
\]
\[\le\limsup\langle
e_2,\tilde\mathcal{T}_{*t}(|f\rangle\langle f|) e_2\rangle^{1/t}\le
e^{a-\eta},\]and since
$\lim\min\sigma(\tilde\mathcal{T}_{*t}(|f\rangle\langle
f|))^{1/t}=e^{a-\eta}$ by (\ref{finite-case-eq1}) we get
 $\lim\langle
e_2,\tilde\mathcal{T}_{*t}(|f\rangle\langle f|)
e_2\rangle^{1/t}=e^{a-\eta}$.
\end{pf}

The existence of an  absorbing pure state can be seen as some
uniform (with respect to the initial state) large deviation
principle, as establishes the following corollary (the implication
$(ii)\Rightarrow (i)$ is a direct consequence of Proposition
\ref{LDP-implies-conv}).

\begin{cor}\label{equiv-LDP}
For any  quantum Markov semigroup $(\mathcal{T}_t)$ on $M_2$ the
following statements are equivalent.
\begin{itemize}
\item[(i)] $(\mathcal{T}_t)$ admits an absorbing pure state;
\item[(ii)] There exists a function $J$ on the  pure state space vanishing at a unique point  such that
for each state $\omega$ distinct from this point,  the net of
orthogonal measures representing $(\mathcal{T}_{*t}(\omega))$
satisfies a large deviation principle with powers $(1/t)$ and  rate
function $J$.
\end{itemize}
When this holds the absorbing state is  the point where $J$
vanishes.
\end{cor}

\begin{rem}\label{remark-NC-LDP}
In \cite{comman-JOT-06}   we defined a noncommutative large
deviation principle for any net of states on any $C^*$-algebra $A$,
where all the basic ingredients of the classical theory are replaced
by their noncommutative counterparts, using the framework of
noncommutative topology. Namely, open (resp. closed) sets are
replaced  by open (resp. closed) projections living in $A^{**}$, and
the  rate function $J$ by a rate operator (more exactly, in order to
avoid possibly infinite-valued operator we use instead the bounded
upper semi-continuous  operator as the counterpart of $\exp-J$,
belonging also to $A^{**}$). Since $A=A^{**}$ when $A$  is finite
dimensional, all self-adjoint operators in $A$ are continuous, in
particular all projections are clopen. In this simple case, by
definition,  a net $(\omega_t)$ of states is said to satisfy a
noncommutative large deviation principle with governing operator $z$
if
\[\lim\omega_t(p)^{1/t}=\sup\{\lambda\in\sigma(z): p
E^z_{\{\lambda\}}\neq 0\} \ \ \ \ \ \textnormal{for all projections
$p\in A$},\] where $\sigma(z)$ and $E^z_{\{\lambda\}}$  denotes
respectively the spectrum
 of $z$ and the eigenspace corresponding to
the eigenvalue $\lambda$ (and $\inf\emptyset=+\infty$ by
convention). The R.H.S. of  the above equality can be written in the
symbolic form $"\sup_p e^{-z}"$  since it is the
 exact noncommutative version of $\sup_{Y} e^{-J}$ for $Y$ open or closed
 (\cite{comman-JOT-06}, Theorem 4.2).
It follows that part $(b)$ of Theorem \ref{equiv-LDP}  amounts to
say
 that
for each state
 $\omega\neq\omega_{|e_1\rangle\langle e_1|}$
 the net of states
$(\mathcal{T}_{*t}(\omega))_{t\ge 0}$ satisfies a
 noncommutative large deviation principle with governing   operator
$\exp{-\mathcal{J}^*(|e_1\rangle\langle e_1|)}$.
\end{rem}

\begin{rem}\label{trivial-LDP}
There are situations not covered by Theorem \ref{finite-case} for
which large deviations, when they hold, are trivial. There are those
 where there is
  a faithful absorbing state $\omega_\infty$; we distinguish two cases.
\begin{itemize}
\item[(a)]   $\omega_\infty\neq\frac{1}{2}I$. By Lemma
\ref{prop-eigen} the hypothesis of Proposition \ref{LDP-converging}
$(a)$ hold, hence (taking $r=2$)
 the associated net of orthogonal measures satisfies a large deviation principle with a
   rate function vanishing at the two
points given by the eigenvectors of $\omega_\infty$, and
infinite-valued elsewhere.
\item[(b)] When $\omega_\infty=\frac{1}{2}I$ the large deviation is
equivalent to the convergence of the one-dimensional projections
given by  eigenvectors, in which case the rate function has the same
form as above. The "if" part of this assertion as well as the form
of the rate function follow from Proposition \ref{LDP-converging}.
Conversely, if a large deviation principle holds for some net of
representing measures, then necessarily the rate function vanishes
on some point, say $\omega_{|e_1\rangle\langle e_1|}$. Varadhan's
theorem implies
\[
\lim\mu_{t}^{1/t}(\widehat{|e_1\rangle\langle e_1|}^t)=\max_{1\le
j\le 2}\{\limsup a_{j,t}^{1/t}|\langle e_1,e_{j,t}\rangle|^2\}\]
\[=\sup_{u\in\mathcal{H},\|u\|=1}|\langle
u,e_1\rangle|^2e^{-J(\omega_{|u\rangle\langle u|})}=1.
\]
Since $\lim a_{1,t}^{1/t}=\lim a_{2,t}^{1/t}=1$ we have
$\lim|\langle e_1,e_{j,t}\rangle|=1$ for some $j$ (say $j=1$) so
that   $\lim |e_{1,t}\rangle\langle e_{1,t}|=|e_1\rangle\langle
e_1|$.
   Since
\[\lim\sum_{j=1}^2 a_{j,t}|\langle
e_1,e_{j,t}\rangle|^2=\omega_\infty(|e_1\rangle\langle
e_1|)=\frac{1}{2}\] with  $\lim a_{1,t}=\lim a_{2,t}=\frac{1}{2}$ we
conclude that $\lim|\langle e_1,e_{2,t}\rangle|=0$ hence $\lim
|e_{2,t}\rangle\langle e_{2,t}|=|e_2\rangle\langle e_2|$ for some
unit vector $e_2$ orthogonal to $e_1$.
\end{itemize}
\end{rem}

\section{Examples}\label{section-ex}

In this section, we study a class of quantum Markov semigroups on
$M_2$ arising in a special instance  of the weak coupling limit (\cite{Fried-48-CMP}, \cite{vanHove_55_Physica},
\cite{Davies-74-CMP}, \cite{Acc-Frig-Lu-90-CMP}). We show that each
of these semigroups has an absorbing state, so that large deviations
follow from Theorem \ref{finite-case} when this state  is pure.
 Before to describe the model we
first recall briefly how works the weak coupling limit theory in
this particular case (\cite{Acc-Lu-91-RMP}, \cite{fa}).

\subsection{Weak coupling limit and squeezed-vacuum
state}\label{weak-coupling}

A quantum system with underlying Hilbert space $\mathcal{H}_0$ is
coupled with the bosonic reservoir on some Hilbert space
$\mathcal{H}_1$ (we shall assume   $\mathcal{H}_1=L^2(\mathbb{R}^d)$
to simplify) where some
  reference
state $\phi$ on the associated CCR algebra  is  given. We will
consider a special case  where $\phi$  is a so-called
\textit{squeezed-vacuum state}, which can be defined as follows. For
each pair of  reals $r,s$ let $T_{r,s}$ be the operator on
$\mathcal{H}_1$ defined by
\[\forall f\in \mathcal{H}_1,\ \ \ \ \ T_{r,s}(f)=(\cosh r)f-\exp(-2is)(\sinh r)\overline{f}.\]
Then $T_{r,s}$ is  real linear, invertible, and so induces a unique
$*$-automorphism of the CCR algebra, defined on the Weyl operators
by $W(f)\mapsto W(T_{r,s}f)$. The vacuum state is transformed by the
above automorphism into the state $\phi_{r,s}$ defined by
\[\forall f\in \mathcal{H}_1,\ \ \ \ \ \phi_{r,s}(W(f))=\exp(-\frac{1}{2}||T_{r,s}f||^2).\]
By means of  the GNS representation of the CCR algebra with state
$\phi_{r,s}$, we obtain for all $f\in \mathcal{H}_1$  a strongly
continuous  unitary group $(W_{r,s}(f))_{t\in\mathbb{R}}$; let
$B_{r,s}(f)$ denote its infinitesimal generator.
 By definition  such a  $\phi_{r,s}$ is a \textit{squeezed-vacuum}
 state if the following conditions hold.
 \begin{itemize}
\item $\phi_{r,s}(B_{r,s}(f))=0$ for all $f\in\mathcal{H}_1$;
\item There exists  $f\in\mathcal{H}_1$ such that
 \[\phi_{r,s}(|B_{r,s}(f)|^2)\neq\phi_{r,s}(|B_{r,s}(if)|^2)\]
 and
 \[\phi_{r,s}(|B_{r,s}(f)|^2)\cdot\phi_{r,s}(|B_{r,s}(if) |^2)
=\frac{1}{4}|\phi_{r,s}([B_{r,s}(f),B_{r,s}(if)])|.\]
 \end{itemize}
In the sequel  we assume that  $\phi$ is such a  squeezed vacuum
state, and  the corresponding $B_{r,s}$ is simply denoted by $B$; we
also fix some $g\in\mathcal{H}_1\verb'\'\{0\}$ and some
$\omega_0>0$. The evolution of the composite system    is given by
the Hamiltonian
\[H^\lambda=H_0\otimes 1+1\otimes H_1+\lambda V,\]
where $H_0$ is the Hamiltonian of the system, $H_1$ is the
Hamiltonian of the free evolution of the reservoir (i.e. the second
quantization of the one-particle Hamiltonian), $\lambda$ the
coupling constant, and
\[V=i(D\otimes  (\frac{1}{2}B(g)-iB(ig))-
D^{\dag}\otimes  (\frac{1}{2}B(g)+iB(ig))),\] where $D$ is a bounded
operator on $H_0$  satisfying
\[\exp(itH_0)D\exp(-itH_0)=\exp(-i\omega_0 t)D\]
and for each $u$ in a dense subset of $\mathcal{H}_0$
\[\sum_{n=1}^{\infty}\frac{|\langle u,D^n
u\rangle|}{[n/2]!}<\infty.\]We put
\[U^\lambda_t=\exp(itH^0)\exp(-itH^\lambda),\ \ \ (t\in\mathbb{R}).\]
We moreover assume that
 the function $t\mapsto\exp(-i\omega_0 t)\langle
f,S_t h\rangle$ is integrable on $\mathbb{R}$ for all $f,h$ in a
dense  linear subspace of the domain of $Q$, where
 $(S_t)$ is the
unitary one-particle free evolution, and
 $Q$ is a real linear operator on $\mathcal{H}_1$ satisfying
 \[\forall f\in\mathcal{H}_1,\ \ \ \ \ \Re\langle f,Q f\rangle=\phi(|B(f)|^2).\]
 Then,
 $U^{\lambda}_{t/\lambda^2}$ converges (in some appropriate sense) to some unitary
transformation $U(t)$ (solution of a quantum stochastic differential
equation) which induces on the algebra $\mathcal{B}(\mathcal{H}_0)$
of all bounded operators on $\mathcal{H}_0$ a quantum Markov
semigroup $(\mathcal{T}_t)$. More precisely, for each normal state
$\omega$ on $\mathcal{B}(\mathcal{H}_0)$ and each
$x\in\mathcal{B}(\mathcal{H}_0)$ we have
\begin{equation}\label{QMS-w-limit}
\lim_{\lambda\rightarrow
0}(\omega\otimes\phi){U^{\lambda}_{t/\lambda^2}}^{\dag}(x\otimes 1)
U^{\lambda}_{t/\lambda^2} =\omega(\mathcal{T}_t(x)).
\end{equation}
  The
generator of $(\mathcal{T}_t)$ is obtained in terms of $D$ and some
parameters depending
 on $\phi$, $g$, $\omega_0$.

\subsection{The model}

We consider a two-level system so that $\mathcal{H}_0=\mathbb{C}_2$,
and we choose
 $D=\left(\begin{array}{cc} 0 & 0 \\
1 & 0
\end{array}\right)$. In this case, the generator $\tilde\mathcal{L}_*$ associated
to the predual semigroup arising as in   (\ref{QMS-w-limit}) has the
Lindblad  form
\[\begin{array}{cl}\forall \rho\in M_2,\ \ \ \ \
\tilde\mathcal{L}_*(\rho)= & i \xi [DD^\dag-D^\dag D,\rho]
 -\frac{\nu+\eta}{2}(D^\dag D \rho-2D\rho D^\dag+\rho D^\dag D)
\\ \\
 & -\frac{\nu}{2}(D D^\dag \rho-2D^\dag\rho
D+\rho D D^\dag) + \overline{\zeta} D\rho D + \zeta D^\dag\rho
D^\dag,
\end{array}
\]
where $\zeta\in\mathbb{C}$ and  $\eta,\nu,\xi$ are reals satisfying
$\eta>0$, $\nu\ge 0$ and
\begin{equation}\label{rel-parameters}
|\zeta|^2\le\nu(\nu+\eta).
\end{equation}
 These parameters depend on the (implicitly fixed) choice of $\phi$, $g$, $\omega_0$, as
described
 in \ref{weak-coupling}; in particular,
 \begin{equation}\label{parameter-eq1}
 \eta=\int_{-\infty}^{+\infty}\exp(-i\omega_0 t)\langle g,S_t g\rangle.
 \end{equation}
Since our large deviation results  will occur when $\nu=0$ (equality
which will be expressed in terms of $\eta$) and the other ones
 will not play any role,
   we do
 not give them here and refer to \cite{fa}.

In order to  show that each of these semigroups has an absorbing
state $\omega_\infty$, we will use the Pauli matrices
$I=\left(\begin{array}{cc} 1 & 0 \\ 0 & 1
\end{array}\right)$, $\sigma_1=\left(\begin{array}{cc} 0 & 1 \\ 1 & 0
\end{array}\right)$, $\sigma_2=\left(\begin{array}{cc} 0 & -i \\ i & 0
\end{array}\right)$, $\sigma_3=\left(\begin{array}{cc} 1 & 0 \\ 0 &
-1\end{array}\right)$. Recall that any self-adjoint operator
$\rho\in M_2$ can be written in a unique way as
$\rho=\frac{1}{2}(\textnormal{tr\ }\rho I+u(\rho)\cdot\sigma)$ where
$u(\rho)=(u_1(\rho),u_2(\rho),u_3(\rho))$ with
$u_i(\rho)=\textnormal{tr}\ \rho\sigma_i$ for $i\in\{1,2,3\}$,
$\sigma=(\sigma_1,\sigma_2,\sigma_3)$, and $u(\rho)\cdot\sigma$
denotes the product $\sum_{i=1}^3u_i(\rho)\sigma_i$. The diagonal
form of $\rho$ is given by
\[\rho=\frac{1}{2}(1+||u(\rho)||)p_{1,\rho}+
\frac{1}{2}(1-||u(\rho)||)p_{2,\rho},\] where
$p_{1,\rho}=\frac{1}{2}(I+\frac{u(\rho)}{||u(\rho)||}\cdot\sigma)$
and
$p_{2,\rho}=\frac{1}{2}(I-\frac{u(\rho)}{||u(\rho)||}\cdot\sigma)$
are the projections on the one-dimensional eigenspaces. Note that
for each real $a$ and each self-adjoint operator $x\in M_2$,
$\textnormal{tr}\frac{1}{2}(I+u(\rho)\cdot\sigma)(aI+u(x)\cdot\sigma)=a+u(\rho)\cdot
u(x)$.

\begin{lem}\label{lemma-rho-t}
\ \ For each positive trace-one operator $\rho\in M_2$ we put
$\rho_t=\tilde{\mathcal{T}}_{*t}(\rho)$ for all $t\ge 0$.
\begin{itemize}
\item[(a)]
 $u_3(\rho_t)=e^{-(2\nu+\eta)t}(u_3(\rho)+\frac{\eta}{2\nu+\eta})-\frac{\eta}{2\nu+\eta}$
\item[(b)] If  $|\zeta|^2-4\xi^2>0$,  then there exist constants
$a_1,a_2,b_1,b_2,c_1,c_2,d_1,d_2$  such that
 \begin{itemize}
 \item $u_1(\rho_t)=e^{m_1t}(a_1u_1(\rho)+b_1u_2(\rho))+e^{m_2t}(a_2u_1(\rho)+b_2u_2(\rho))$.
 \item $u_2(\rho_t)=e^{m_1t}(c_1u_1(\rho)+d_1u_2(\rho))+e^{m_2t}(c_2u_1(\rho)+d_2u_2(\rho))$.
 \end{itemize}
 where $m_1=-(\nu+\frac{\eta}{2})+\sqrt{|\zeta|^2-4\xi^2}$,
 $m_2=-(\nu+\frac{\eta}{2})-\sqrt{|\zeta|^2-4\xi^2}$.
 \item[(c)] If $|\zeta|^2-4\xi^2<0$, then there exist constants
$a_1,a_2,b_1,b_2,c_1,c_2,d_1,d_2$ such that
\begin{itemize}
 \item $u_1(\rho_t)=e^{mt}((a_1u_1(\rho)+b_1u_2(\rho))\cos
 t\sqrt{4\xi^2-|\zeta|^2})+\\(a_2u_1(\rho)+b_2u_2(\rho))\sin
 t\sqrt{4\xi^2-|\zeta|^2})$
 \item $u_2(\rho_t)=e^{mt}((c_1u_1(\rho)+d_1u_2(\rho))\cos
 t\sqrt{4\xi^2-|\zeta|^2})+\\(c_2u_1(\rho)+d_2u_2(\rho))\sin
 t\sqrt{4\xi^2-|\zeta|^2})$
 \end{itemize}
 where $m=-(\nu+\frac{\eta}{2})$.
\item[(d)] If $|\zeta|^2-4\xi^2=0$, then there exist constants
$a_1,a_2,b_1,b_2,c_1,c_2,d_1,d_2$ such that
\begin{itemize}
\item $u_1(\rho_t)=e^{2mt}((a_1u_1(\rho)+b_1u_2(\rho))+t(a_2u_1(\rho)+b_2u_2(\rho)))$.
\item $u_2(\rho_t)=e^{2mt}((c_1u_1(\rho)+d_1u_2(\rho))+t(c_2u_1(\rho)+d_2u_2(\rho)))$.
\end{itemize}
where $m=-(\nu+\frac{\eta}{2})$.
 \end{itemize}
\end{lem}

\begin{pf}
Direct calculations yield
\begin{itemize}
\item[(i)] $\tilde\mathcal{L_*}(I)=-\eta\sigma_3$
\item[(ii)]
$\tilde\mathcal{L_*}(\sigma_1)=-(\nu+\frac{\eta}{2}-\Re\zeta)\sigma_1+
(2\xi-\Im\zeta)\sigma_2$,
\item[(iii)] $\tilde\mathcal{L_*}(\sigma_2)=-(2\xi+\Im\zeta)\sigma_1-(\nu+\frac{\eta}{2}+
\Re\zeta)\sigma_2$,
\item[(iv)] $\tilde\mathcal{L_*}(\sigma_3)=-(2\nu+\eta)\sigma_3$.
\end{itemize}
Differentiating  $\tilde{\mathcal{T}}_{*t}=e^{t\tilde\mathcal{L_*}}$
yields
$\frac{d\tilde{\mathcal{T}}_{*t}}{dt}=\tilde\mathcal{L_*}\circ\tilde{\mathcal{T}}_{*t}$,
hence the following system of differential equations
\begin{itemize}
\item $\frac{d\tilde{\mathcal{T}}_{*t}(I)}{dt}=-\eta \tilde{\mathcal{T}}_{*t}(\sigma_3)$
\item
$\frac{d\tilde{\mathcal{T}}_{*t}(\sigma_1)}{dt}=-(\nu+\frac{\eta}{2}-\Re\zeta)
\tilde{\mathcal{T}}_{*t}(\sigma_1)+(2\xi-\Im\zeta)\tilde{\mathcal{T}}_{*t}(\sigma_2)$
\item $\frac{d\tilde{\mathcal{T}}_{*t}(\sigma_2)}{dt}=-(2\xi+\Im\zeta)\tilde{\mathcal{T}}_{*t}(\sigma_1)-(\nu+\frac{\eta}{2}+
\Re\zeta)\tilde{\mathcal{T}}_{*t}(\sigma_2)$
\item $\frac{d\tilde{\mathcal{T}}_{*t}(\sigma_3)}{dt}=-(2\nu+\eta)\tilde{\mathcal{T}}_{*t}(\sigma_3)$.
\end{itemize}
with initial conditions $\tilde{\mathcal{T}}_{*0}(I)=I$,
$\tilde{\mathcal{T}}_{*0}(\sigma_1)=\sigma_1$,
$\tilde{\mathcal{T}}_{*0}(\sigma_2)=\sigma_2$,
$\tilde{\mathcal{T}}_{*0}(\sigma_3)=\sigma_3$. We then obtain
\[
\widetilde{\mathcal{T}}_{*t}(I)=I+\frac{\eta}{2\nu+\eta}(e^{-(2\nu+\eta)t}-1)\sigma_3.
\]
\[
\widetilde{\mathcal{T}}_{*t}(\sigma_3)=e^{-(2\nu+\eta)t}\sigma_3,
\]
which gives $(a)$. Put $x(t)=\tilde{\mathcal{T}}_{*t}(\sigma_1)$,
$y(t)=\tilde{\mathcal{T}}_{*t}(\sigma_2)$. Clearly, $x(t)$  has only
non-zero components $x_1(t)$  on $\sigma_1$ and $x_2(t)$ on
$\sigma_2$; similarly, $y(t)=(y_1(t),y_2(t))$.
 The characteristic
equation associated to the matrix given by the system
\begin{equation}\label{lemma-rho-t-eq0}
 \left\{
 \begin{array}{ll}
x'(t)=-(\nu+\frac{\eta}{2}-\Re\zeta)x(t)
+(2\xi-\Im\zeta)y(t) & \\
\\
y'(t)=-(2\xi+\Im\zeta)x(t)-(\nu+\frac{\eta}{2}+ \Re\zeta)y(t) &
\end{array}
\right.
\end{equation}
with initial conditions $x(0)=(\sigma_1,0)$, $y(0)=(0,\sigma_2)$,
 is on each component
 \begin{equation}\label{lemma-rho-t-eq1}
 X^2+(2\nu+\eta)X+\frac{\eta^2}{2}+\nu^2+\nu\eta+4\xi^2-|\zeta|^2\end{equation}
 and
 so $\Delta=4(|\zeta|^2-4\xi^2)$.

 \bigskip

Assume  $|\zeta|^2-4\xi^2>0$. The solutions of
(\ref{lemma-rho-t-eq1}) being
$m_1=-(\nu+\frac{\eta}{2})+\sqrt{|\zeta|^2-4\xi^2}$ and
 $m_2=-(\nu+\frac{\eta}{2})-\sqrt{|\zeta|^2-4\xi^2}$,
 the  general solution of (\ref{lemma-rho-t-eq0}) is
 \[
 \left\{
 \begin{array}{ll}
 x_1(t)=a_1e^{m_1t}+a_2e^{m_2t}, &  x_2(t)=c_1e^{m_1t}+c_2e^{m_2t} \\
\\
y_1(t)=b_1e^{m_1t}+b_2e^{m_2t}, & y_2(t)=d_1e^{m_1t}+d_2e^{m_2t}
\end{array}
\right.
\]
for suitable constants $a_1,a_2,b_1,b_2,c_1,c_2,d_1,d_2$. It follows
that
\[
 \tilde{\mathcal{T}}_{*t}(\sigma_1)=(a_1e^{m_1t}+a_2e^{m_2t})\sigma_1+
 (c_1e^{m_1t}+c_2e^{m_2t})\sigma_2,
 \]
 \[
 \tilde{\mathcal{T}}_{*t}(\sigma_2)=(b_1e^{m_1t}+b_2e^{m_2t})\sigma_1
+(d_1e^{m_1t}+d_2e^{m_2t})\sigma_2,
 \]
 which gives $(b)$.

 \bigskip

Assume
 $|\zeta|^2-4\xi^2<0$. Then
$m_1=-(\nu+\frac{\eta}{2})+i\sqrt{4\xi^2-|\zeta|^2}$,
 $m_2=-(\nu+\frac{\eta}{2})-i\sqrt{4\xi^2-|\zeta|^2}$, and the
 general solution of  (\ref{lemma-rho-t-eq0}) is
\[
 \left\{
 \begin{array}{ll}
 x_1(t)=e^{mt}(a_1\cos t\sqrt{4\xi^2-|\zeta|^2}+a_2\sin
 t\sqrt{4\xi^2-|\zeta|^2})\\
\\
y_1(t)=e^{mt}(b_1\cos\sqrt{4\xi^2-|\zeta|^2}+b_2\sin\sqrt{4\xi^2-|\zeta|^2})
\end{array}
\right.
\]
\[
 \left\{
 \begin{array}{ll}
  x_2(t)=e^{mt}(c_1\cos t\sqrt{4\xi^2-|\zeta|^2}+c_2\sin t\sqrt{4\xi^2-|\zeta|^2})\\
\\
y_2(t)=e^{mt}(d_1\cos\sqrt{4\xi^2-|\zeta|^2}+d_2\sin\sqrt{4\xi^2-|\zeta|^2})
\end{array}
\right.
\]
 where $m=-(\nu+\frac{\eta}{2})$, and
$a_1,a_2,b_1,b_2,c_1,c_2,d_1,d_2$ are suitable constants. We then
obtain
 \[
 \tilde{\mathcal{T}}_{*t}(\sigma_1)=e^{mt}(a_1\cos t\sqrt{4\xi^2-|\zeta|^2}+a_2\sin
 t\sqrt{4\xi^2-|\zeta|^2})\sigma_1+
\]
 \[e^{mt}(c_1\cos t\sqrt{4\xi^2-|\zeta|^2}+c_2\sin
 t\sqrt{4\xi^2-|\zeta|^2})\sigma_2.\]
 \[
\tilde{\mathcal{T}}_{*t}(\sigma_2)=e^{mt}(b_1\cos
t\sqrt{4\xi^2-|\zeta|^2}+b_2\sin
 t\sqrt{4\xi^2-|\zeta|^2})\sigma_1+
 \]
 \[e^{mt}(d_1\cos t\sqrt{4\xi^2-|\zeta|^2}+d_2\sin
 t\sqrt{4\xi^2-|\zeta|^2})\sigma_2,\] and $(c)$ follows
 easily.

\bigskip

If
 $|\zeta|^2-4\xi^2=0$, then the unique solution of
(\ref{lemma-rho-t-eq1}) is $2m$  with $m=-(\nu+\frac{\eta}{2})$, and
the general  solution of (\ref{lemma-rho-t-eq0}) is
\[
 \left\{
 \begin{array}{ll}
 x_1(t)=e^{2mt}(a_1+a_2t), &  x_2(t)=e^{mt}(c_1+c_2t)\\
\\
 y_1(t)=e^{2mt}(b_1+b_2t), &  y_2(t)=e^{mt}(d_1+d_2t)\\
\end{array}
\right.
\]
for suitable constants $a_1,a_2,b_1,b_2,c_1,c_2,d_1,d_2$, so that
\[\tilde{\mathcal{T}}_{*t}(\sigma_1)=e^{2mt}(a_1+a_2t)\sigma_1+e^{2mt}(c_1+c_2t)\sigma_2\]
\[\tilde{\mathcal{T}}_{*t}(\sigma_2)=e^{2mt}(b_1+b_2t)\sigma_1+e^{2mt}(d_1+d_2t)\sigma_2,\]
which imply $(d)$.
\end{pf}

\begin{prop}\label{absorb-state}
The state $\omega_\infty$ given by the operator
$\rho_\infty=\frac{1}{2}(I-\frac{\eta}{(2\nu+\eta)}\sigma_3)$ is
absorbing for the semigroup $(\mathcal{T}_t)_{t\ge 0}$.
\end{prop}

\begin{pf}
Let $\omega_\rho$  be a state  and put
$\rho_t=\tilde{\mathcal{T}}_{*t}(\rho)$ for all $t\ge 0$.  The
hypotheses $\eta>0$, $|\zeta|^2\le\nu(\nu+\eta)$, $\nu\ge 0$ imply
(with the notations of Lemma \ref{lemma-rho-t})
 $m_1m_2=\frac{\eta^2}{2}+\nu^2+\nu\eta+4\xi^2-|\zeta|^2>0$, and
 so
 $m_1<0$, $m_2< 0$, $m<0$. It follows that $\lim u_1(\rho_t)=0=\lim u_2(\rho_t)$
 and $\lim u_3(\rho_t)=\frac{-\eta}{(2\nu+\eta)}$, hence for any
 $\rho$ initial, and any positive
 $x=aI+u(x)\cdot\sigma$ in $M_2$, we have
 \[\lim\textnormal{tr}\rho_t x=\lim(a+u(\rho_t)\cdot u(x))=
 (a-\frac{\eta}{2\nu+\eta}u_3(x))=\textnormal{tr}\rho_\infty
 x.\]
\end{pf}

The above proposition shows that  $\omega_\infty$ is pure if and
only if $\nu=0$. As it is easily seen from the expressions given in
\cite{fa}, we have
\begin{equation}\label{parameters-eq3}
\nu=0\ \ \ \Longleftrightarrow\ \ \ \eta=\frac{1}{2}
 \Re\int_{-\infty}^{+\infty}\exp(-i\omega_0 t)(\langle g,S_t Q(g)\rangle
 +\langle ig,S_t Q(ig)\rangle).
 \end{equation}
Note that this holds in particular when $Q=I$ (i.e. $\phi$ is the
vacuum
 state).
We  have $\rho_\infty=\left(\begin{array}{cc} 0\ & 0
\\ 0 & 1
\end{array}\right)(=|e_1\rangle\langle e_1|)$,  and since $\zeta=0$ by
(\ref{rel-parameters}),  the Lindblad form of the generator becomes
\[\rho\in M_2,\ \ \ \ \
\tilde\mathcal{L}_*(\rho)= i \xi [DD^\dag -D^\dag D,\rho]
 -\frac{\eta}{2}(D^\dag D \rho-2D\rho D^\dag+\rho D^\dag  D),
\]
or equivalently
  as in Theorem \ref{Davies-result},
\[\forall \rho\in M_2,\ \ \ \ \ \ \tilde\mathcal{L}_*(\rho)=y\rho+\rho y^*+z\rho z^*,\]
with $y=\left(\begin{array}{cc} -\frac{\eta}{2}-i \xi\ & 0 \\
0 & i\xi
\end{array}\right)$ and $z=\left(\begin{array}{cc} 0 & 0 \\ \sqrt\eta &
0
\end{array}\right)$; in particular $\mathcal{J}^*(|e_1\rangle\langle e_1|)=
\eta|e_2\rangle\langle e_2|$.

 The following large deviation result
 follows from Theorem \ref{finite-case} and Corollary \ref{NC-LDP}. It shows
that the exponential asymptotic behavior of $(\mathcal{T}_{*t})$ is
controlled by the parameter (\ref{parameter-eq1}); moreover, it does
not depend on the choice of the  squeezed-vacuum state $\phi$,
provided that $\phi$ satisfies the condition of
(\ref{parameters-eq3}).

\begin{prop}\label{LDP-first-case}
 If $\nu=0$, then for each initial state $\omega\neq\omega_{\rho_\infty}$
 the following conclusions hold.
\begin{itemize}
\item[(a)]
 The net of orthogonal measures representing
$({\mathcal{T}}_{*t}(\omega))_{t\ge 0}$ satisfies a large deviation
principle with powers $(1/t)$ and rate function
\begin{displaymath}
 J(\omega_{|e\rangle\langle
e|})=\left\{
\begin{array}{ll}
0 & \ \ \ \ \ \ \textnormal{if $|e\rangle\langle e|=\rho_\infty$}
\\
\eta & \ \ \ \ \ \ \textnormal{if $|e\rangle\langle
e|=I-\rho_\infty$}
\\
+\infty & \ \ \ \ \ \ otherwise.
\end{array}
\right.
\end{displaymath}
\item[(b)] For  each  projection $p\in
M_2\verb'\'\{0\}$ we have
\[
\lim\frac{1}{t}\log\omega(\mathcal{T}_t (p))=\left\{
\begin{array}{ll}
-\eta & \ \ \ \ \ \ \textnormal{if $p=I-\rho_\infty$}
\\ \\
0 & \ \ \ \ \ \ otherwise.
\end{array}
\right.
\]
\end{itemize}
\end{prop}

\begin{rem}\label{second-proof}
The explicit expressions of ${\mathcal{T}}_{*t}(\omega)$ given by
Lemma \ref{lemma-rho-t} $(c)$   allow a direct proof of Proposition
\ref{LDP-first-case} $(a)$. Indeed, easy calculations yield
 \[u_1(\rho_t)=e^{-\frac{\eta}{2}t}(u_1(\rho)\cos(2\xi t)-u_2(\rho)\sin(2\xi
t)),\]
\[u_2(\rho_t)=e^{-\frac{\eta}{2}t}(u_1(\rho)\sin(2\xi
t)+u_2(\rho)\cos(2\xi t)),\]so that
\[1-||u(\rho_t)||^2=-e^{-2\eta t}(1+u_3(\rho))^2-
e^{-\eta t}(u_1(\rho)^2+u_2(\rho)^2-2(u_3(\rho)+1)).\]If
$u_1(\rho)^2+u_2(\rho)^2-2(u_3(\rho)+1)=0$, then necessarily
$u_3(\rho)=-1$, $u_1(\rho)=u_2(\rho)=0$, and $\rho=\rho_\infty$,
which is excluded. It follows that
$u_1(\rho)^2+u_2(\rho)^2-2(u_3(\rho)+1)<0$ and
$\lim(1-||u(\rho_t)||^2)^{1/t}=e^{-\eta}$. Since
 $\lim
(1+||u(\rho_t)||)^{1/t}=1$ and
\[e^{-\eta}=\lim(1-||u(\rho_t)||^2)^{1/t}=\limsup(1-||u(\rho_t)||)^{1/t}
\lim(1+||u(\rho_t)||)^{1/t},\] we get
 $\lim(1-||u(\rho_t)||)^{1/t}=e^{-\eta}$.  Since
$\rho_t=\frac{1}{2}(1+||u(\rho_t)||)p_{1,\rho_t}+
\frac{1}{2}(1-||u(\rho_t)||)p_{2,\rho_t}$ the conclusion follows
from Proposition \ref{LDP-converging}.
\end{rem}

\section*{Acknowledgments}
The author wishes  to thank S. Attal for many stimulating
discussions, as well as  for the support and the warm hospitality he
enjoyed during a visit at the Institut Jordan  in 2006. This work
has been supported by FONDECYT grant No. 7070061.

\pagebreak

\end{document}